\documentclass[12pt,a4paper]{article}
\usepackage{amsmath,amssymb,epsfig}
\newcommand{\beq}{\begin{eqnarray}}
\newcommand{\eeq}{\end{eqnarray}}

\newcommand{\be}{\begin{equation}}
\newcommand{\ee}{\end{equation}}

\newcommand{\bm}{\begin{multline}}
\newcommand{\fm}{\end{multline}}

\begin{document}
\setlength{\unitlength}{.8mm}

\begin{titlepage} 
\vspace*{0.5cm}
\begin{center}
{\Large\bf Finite size effects and 2-string deviations in the spin-1 XXZ chains}
\end{center}
\vspace{1.5cm}
\begin{center}
{\large \'Arp\'ad Heged\H us}
\end{center}
\bigskip

\vspace{0.1cm}

\begin{center}
Research Institute for Particle and Nuclear Physics,\\
Hungarian Academy of Sciences,\\
H-1525 Budapest 114, P.O.B. 49, Hungary\\ 
\end{center}
\vspace{2.5cm}
\begin{abstract}

In an earlier work \cite{1} J. Suzuki proposed a set of nonlinear integral equations (NLIE) to describe the 
excited state spectrum of the integrable spin-1 XXZ chain in its repulsive regime. 
In this paper we extend his 
equations for the attractive regime of the model, and calculate analytically the conformal spectrum of the 
spin chain. We 
also discuss the typical root configurations of the thermodynamic limit as well as the 2-string deviations 
of certain excited states of the model. Special objects appearing in the NLIE are also treated with special 
care.

\end{abstract}

\end{titlepage}

\section{Introduction}

Recently spin-1 chains and quantum field theories related to the 19-vertex model \cite{2} attract much 
attention. The spin-1 XXZ chains deducible from the 19-vertex model are interesting because on the one hand
 their Hamiltonian occurs in large $N_c$ QCD as the 1-loop anomalous dimension 
matrix of the single trace operators containing the self-dual components of the 
field strength tensor \cite{3}, and on the other hand the determination of correlation functions in these
integrable higher spin chains is still an area of active research \cite{4}.
Furthermore it is known that in the context of light-cone approach \cite{5} 
the inhomogeneous 19-vertex model with alternating inhomogeneities provides an integrable lattice
regularization for the ${\cal N}=1$ supersymmetric sine-Gordon model \cite{6,7}.     

In this paper we investigate the finite size effects and 2-string deviations of 
the integrable spin-1 XXZ chain.
It is well known that in the thermodynamic limit the 
anti-ferromagnetic ground state of the model is composed of quasi 2-strings, namely of pairs of complex 
roots
 having imaginary parts close to $\pm \frac{\pi}{2}$. The deviations of their imaginary parts from 
 the values $\pm \frac{\pi}{2}$ are called 2-string deviations. According to the string hypothesis
 these deviations should be exponentially small in $N$ \cite{8,9,10}, but it turns out that these deviations
are much larger; they are of order $1/N$ \cite{11,12}. 
This is because the number of 2-strings is not of order one,
 but $N/2$ minus $O(1)$ such that the centers of the outermost 2-strings tend to infinity as $N$
 tends to infinity.  This is why these 2-string deviations are not negligible even in the
large $N$ limit calculations of the physical quantities.
To treat correctly the technical difficulties coming from 2-string deviations one needs to use 
the so-called nonlinear integral equation technique (NLIE).
The NLIE technique was originally introduced in \cite{12} where with the help of this technique the
finite size effects of the ground states of the spin-1/2 and spin-1 XXZ chains were studied calculating
analytically their central charges and in the spin-1 case the 2-string deviations in the ground state
 for the first time in a unified NLIE approach. 
 Then the NLIE technique was successfully applied for describing finite size effects in 
various integrable spin chains and quantum field theories \cite{13,14,15,16,17,18,19}.

Recently J. Suzuki \cite{1} derived a set of NLIEs (different from the those of \cite{12}) 
to describe the excited states of the integrable spin-1 XXZ chain 
in its repulsive regime (i.e. $0<\gamma<\frac{\pi}{3}$). In this paper we extend his results also for the
attractive regime of the model (i.e. $\frac{\pi}{3}<\gamma<\frac{\pi}{2}$). 
With the help of our NLIEs, we determine the operator content of the conformal field theory 
describing the thermodynamic limit of the spin chain. 
Furthermore the NLIE enables us to discuss the typical root configurations and 2-string deviations of 
excited states in the large $N$ limit.

The paper is organized as follows: 
In section 2 some generalities concerning Bethe Ansatz solution of the models under consideration are 
recalled. In section 3 the classification of Bethe roots is presented. The auxiliary functions
and their most important properties are listed in section 4. The presentation of the NLIEs can be found in 
section 5. Section 6 contains the counting equations, and section 7 is devoted for the discussion
of typical root configurations of the thermodynamic limit. Section 8 contains the quantitative
analysis of 2-string deviations and in section 9 the calculation of the conformal spectrum of the spin 
chain is presented. Section 10 is devoted to special objects.
Finally the summary and perspectives for future work can be found in section 11.

\section{The definition of the model}

The integrable spin-1 XXZ chain is defined by its Hamiltonian:

\begin{equation} \label{H}
{\cal H}=\sum_{i=1}^N \left( \sigma_i^\bot-(\sigma_i^\bot)^2+\cos 2\gamma \,
(\sigma_i^z-(\sigma_i^z)^2)-(2  \, \cos \gamma -1)(\sigma_i^\bot \sigma_i^z+ \sigma_i^z \sigma_i^\bot)-4 
\sin^2 \gamma (S_i^z)^2
 \right),
\end{equation}
where $$\sigma_i=S_i\cdot S_{i+1}=\sigma_i^\bot+\sigma_i^z, \qquad \sigma_i^z=S_i^z\cdot S_{i+1}^z.$$
In the rest of the paper we consider N to be even and we impose periodic boundary conditions on the spins 
$S_{N+1}^a=S_1^a, \, a \in \{x,y,z \}$.

It is well known that the above spin Hamiltonian can be obtained from the transfer matrix of the 19-vertex 
model \cite{8,9,10}. Hereafter we consider the 19-vertex model with alternating inhomogeneities in order 
to get access to the light-cone Hamiltonian of the ${\cal N}=1$ supersymmetric sine-Gordon model as well.

Let $V_i \simeq {\mathbb C}^{l_i+1} $ be the irreducible $SU(2)$ representation with spin $l_i/2$ and let 
$R_{ij}^{(l_i,l_j)}(\theta)$ the $R$-matrices acting on $V_i \otimes V_j$ satisfying the Yang-Baxter 
equations
\begin{equation} \label{YB}
R_{12}^{(l_1,l_2)}(\theta) R_{13}^{(l_1,l_3)}(\theta+\theta') R_{23}^{(l_2,l_3)}(\theta')=
R_{23}^{(l_2,l_3)}(\theta') R_{13}^{(l_1,l_3)}(\theta+\theta') R_{12}^{(l_1,l_2)}(\theta).
\end{equation}
These $R$-matrices can be obtained by fusion \cite{20} from the well known $R$-matrix of the six-vertex 
model, and their explicit form can be found in \cite{9,10}.
 
 From these $R$-matrices one can define a family of transfer matrices with alternating inhomogeneities 
$\lambda_i=(-1)^i \Theta$:
\begin{equation} \label{TM}
T_k(\theta,\{\lambda_i \})=\mbox{Tr}_a \left(R_{a1}^{(k,2)}(\theta-\lambda_1-i \frac{\pi}{2}(1+k))\dots 
R_{aN}^{(k,2)}(\theta-\lambda_N-i \frac{\pi}{2}(1+k)) \right). 
\end{equation}
Due to the Yang-Baxter relation (\ref{YB}) the transfer matrices (\ref{TM}) form a commutative family
of operators acting on $V_H=\otimes_{i=1}^N \, {\mathbb C}^3$:
\begin{equation}
\big[ T_k(\theta,\{\lambda_i \}),T_{k'}(\theta',\{\lambda_i \}) \big]=0.
\end{equation}
Investigating only the case of alternating inhomogeneities we can write for short 
$T_k(\theta,\Theta)=T_k(\theta,\{(-1)^i \Theta \})$.
Due to integrability and commutativity all these transfer matrices can be diagonalized by 
algebraic Bethe Ansatz \cite{27}. The eigenvalues of the transfer matrices can be characterized by the
solutions of the Bethe Ansatz equations:
\begin{equation} \label{BAE}
\frac{\Phi(\theta_j+i \pi)}{\Phi(\theta_j-i \pi)}=-\frac{Q(\theta_j+i\pi)}{Q(\theta_j-i\pi)}, \qquad
j=1,\dots,M,
\end{equation}
where
\begin{equation} \label{}
\Phi(\theta)=\left( \sinh\frac{\gamma}{\pi}(\theta-\Theta)\sinh\frac{\gamma}{\pi}(\theta+\Theta) \right)^{N/2},
\end{equation}
\begin{equation} \label{}
Q(\theta)=\prod_{j=1}^{M} \sinh\frac{\gamma}{\pi}(\theta-\theta_j),
\end{equation}
where $\gamma$ is the anisotropy of the model, $M$ is the number of Bethe roots.
We recall that $S=N-M$ is the third component of the total spin of the spin chain.
The eigenvalues of the first two transfer matrices of the fusion hierarchy can be expressed by the solutions of 
the Bethe Ansatz equations as follows:
\begin{equation} \label{T1}
T_1(\theta,\Theta)=\Phi(\theta-i\pi) \frac{Q(\theta+i\pi)}{Q(\theta)}+\Phi(\theta+i\pi) 
\frac{Q(\theta-i\pi)}{Q(\theta)},
\end{equation}
\begin{eqnarray} \label{T2}
T_2(\theta,\Theta) &=& \Phi(\theta-i\frac{\pi}{2})\Phi(\theta-i\frac{3\pi}{2})
 \frac{Q(\theta+i\frac{3\pi}{2})}{Q(\theta-i\frac{\pi}{2})}+\Phi(\theta+i\frac{\pi}{2})\Phi(\theta+i\frac{3
\pi}{2})
\frac{Q(\theta-i\frac{3\pi}{2})}{Q(\theta+i\frac{\pi}{2})} \nonumber \\
&+& \Phi(\theta-i\frac{\pi}{2})\Phi(\theta+i\frac{\pi}{2})\frac{Q(\theta+i\frac{3\pi}{2})}{Q(\theta-
i\frac{\pi}{2})} \frac{Q(\theta-i\frac{3\pi}{2})}{Q(\theta+i\frac{\pi}{2})}.
\end{eqnarray}
The spin Hamiltonian (\ref{H}) is given by the logarithmic derivative of the homogeneous transfer matrix 
$T_2(\theta,0)$:
\begin{equation} \label{HT2}
{\cal H}=\frac{\pi}{\gamma} \frac{d}{d\theta} \, \log \, T_2(\theta,0)|_{\theta=-i \frac{\pi}{2}}.
\end{equation} 
On the other hand the regularized finite volume light-cone hamiltonian of the ${\cal N}=1$ 
supersymmetric sine-Gordon model
can be expressed by the inhomogeneous transfer matrix  $T_2(\theta,\Theta):$
\begin{equation} \label{EPM}
e^{i a (H\pm P)/2} \sim T_2(\pm \Theta \pm i \frac{\pi}{2},\Theta),
\end{equation}
where $a=L/N$ is the lattice constant, and the continuum limit is achieved by taking $N$ to infinity
along with tuning the inhomogeneity parameter as $\Theta=\ln\frac{2N}{mL}$ with $m$ being the kink mass.

\section{Classification of Bethe roots}

First of all let us parametrize the anisotropy $\gamma$ as
\begin{equation} \label{gp}
\gamma=\frac{\pi}{p+2}, \qquad 0<p.
\end{equation}
In this notation the attractive ($\frac{\pi}{3}<\gamma<\frac{\pi}{2}$) and the repulsive 
($0<\gamma<\frac{\pi}{3}$) regimes of the model correspond to the $0<p<1$ and $1<p$ regions
 respectively.

All functions entering the Bethe Ansatz equations (\ref{BAE}) are periodic with respect to $\pi(p+2)$, 
thus on the complex plane the domain 
of Bethe roots can be restricted to the strip $-\frac{\pi(p+2)}{2}<\mbox{Im}\theta\leq \frac{\pi(p+2)}{2}.$
The Bethe roots are either real or come in complex conjugated pairs, with the exception of self-conjugated 
roots
with $\mbox{Im} \, \theta_j=\frac{\pi(p+2)}{2}$. It is well known that the ground state of the model is
 formed by $N/2$ quasi 2-strings, namely pairs of Bethe roots with imaginary parts being close to $\pm 
\frac{\pi}{2}$ \cite{8,9,10}. The root configurations of the excitations can be obtained from that of 
the ground state by adding some other new Bethe roots and removing some 2-strings. 
 We classify the  Bethe roots as follows:  (See figure 1.)\newline
\newline
1. \emph{Inner roots}: $|\mbox{Im} \, \theta_j|< \frac{\pi}{2},$ 
\newline
\newline 
2. \emph{Close roots}: $\frac{\pi}{2}<|\mbox{Im} \, \theta_j|< \frac{\pi}{2}+\mbox{min}(1,p) \, \pi,$ 
\newline
\newline 
3. \emph{Wide roots}: $\frac{\pi}{2}+\mbox{min}(1,p) \, \pi<|\mbox{Im} \, \theta_j|< \frac{\pi(p+2)}{2},$ 
\newline
\newline 
4. \emph{Self-conjugated roots}: $|\mbox{Im} \, \theta_j|= \frac{\pi(p+2)}{2}, $
\newline

In this classification the quasi 2-strings can be either inner- or close-roots. Having derived the NLIE
governing the finite size effects of the model we will give a more precise classification of the Bethe roots.

It is worth introducing the concept of \emph{effective roots} as well.
To each root $\theta_j$ with $\frac{\pi}{2}<|\mbox{Im} \theta_j|\leq\frac{\pi(p+2)}{2}$ we associate
an \emph{"effective root"} $\tilde{\theta}_j$ by the following definition:
\begin{equation} \label{effdef}
\tilde{\theta}_j=\theta_j-i \frac{\pi}{2}  \mbox{sign}(\mbox{Im} \, \theta_j).
\end{equation}
This transformation is equivalent to the removal of the middle strip of inner roots from the fundamental
domain of the original Bethe roots. (See figure 2.)
\begin{figure}[htb]
\begin{flushleft}
\hskip 15mm
\leavevmode
\epsfxsize=160mm
\epsfbox{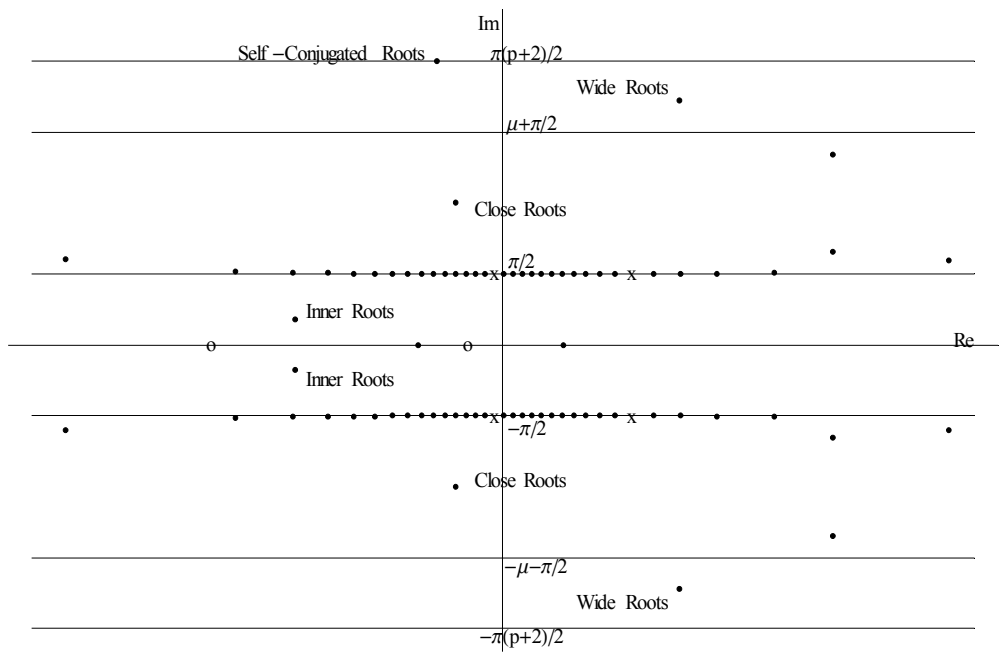}
\end{flushleft}
\caption{{\footnotesize
Classification of Bethe roots. Crosses stand for the 2-string holes, empty circles 
denote the type I holes, black dots represent the Bethe roots and $\mu$ denotes $\pi \mbox{min}(1,p)$. 
}}
\label{1}
\end{figure}

The classification of \emph{effective roots} is as follows
\newline 
1. \emph{Close effective roots}: $0<|\mbox{Im} \, \tilde{\theta_j}|< \mbox{min}(1,p) \, \pi,$ 
\newline
\newline 
2. \emph{Wide effective roots}: $\mbox{min}(1,p) \, \pi<|\mbox{Im} \, \tilde{\theta_j}|< \frac{\pi(p+1)}{2},$ 
\newline
\newline
3. \emph{Self-conjugated effective roots}: $|\mbox{Im} \, \tilde{\theta_j}|= \frac{\pi(p+1)}{2}, $
\newline
\begin{figure}[htb]
\begin{flushleft}
\hskip 15mm
\leavevmode
\epsfxsize=160mm
\epsfbox{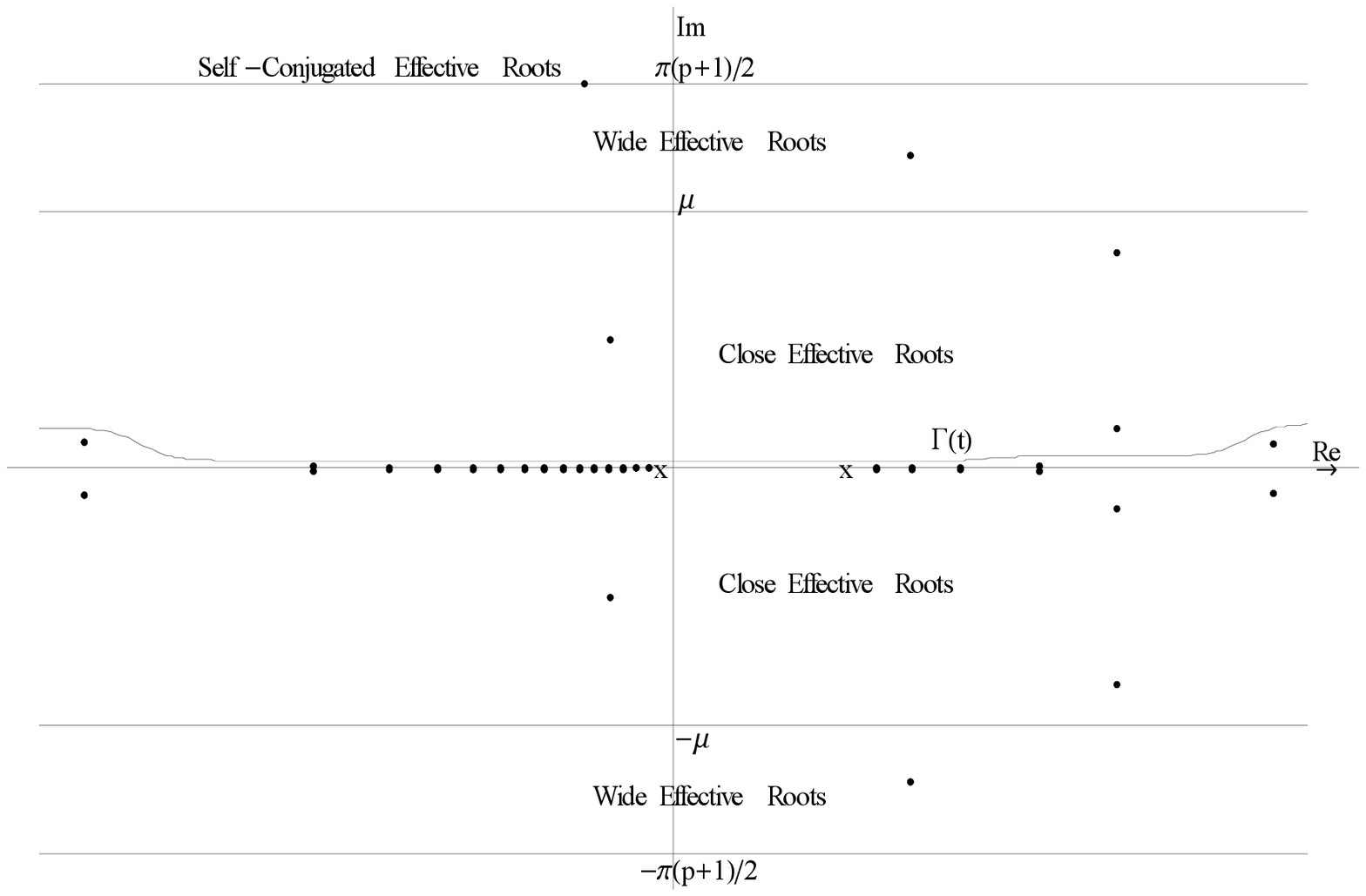}
\end{flushleft}
\caption{{\footnotesize
Classification of effective roots. Crosses stand for holes, effective roots are represented by
 black dots  and $\mu$ denotes $\pi \mbox{min}(1,p)$. 
}}
\label{2}
\end{figure}

The fundamental domain and the classification of effective roots are exactly the same as those of the
Bethe roots in the 6-vertex model, corresponding to the bosonic degrees of freedom of the
conformal field theory describing the thermodynamic limit of the spin chain.

\section{Auxiliary functions and their most important properties}

In this paper we study the finite size effects of our model with the help of a set of nonlinear integral 
equations (NLIE). In this technique the contribution of the $N/2$ minus $O(1)$ number of
 quasi 2-strings is summed up by
the integral terms of the NLIE, and one has to deal with only a finite number of holes and other type 
of complex roots characterizing the excitation \cite{1}. This technique is very efficient when one is 
interested in the spectrum of our model in the thermodynamic limit.
 The first step to formulate the NLIE is the definition of proper auxiliary functions. Here we use
the auxiliary functions introduced in \cite{1,15}.

The most natural auxiliary function $a(\theta)$ is defined by
\begin{equation} \label{a}
a(\theta)=\frac{\Phi(\theta+i \pi)}{\Phi(\theta-i \pi)}\frac{Q(\theta-i \pi)}{Q(\theta+i \pi)},
\end{equation}
\begin{equation} \label{1+a}
1+a(\theta)=\frac{1}{\Phi(\theta-i \pi)}\frac{Q(\theta)}{Q(\theta+i \pi)} T_1(\theta).
\end{equation}
By the help of $a(\theta)$ the Bethe Ansatz equations (\ref{BAE}) can be recasted into the form
$$a(\theta_j)=-1, \qquad j=1,\dots,M.$$ Written in logarithmic form
$$i \, \log a(\theta_j)=2 \pi I_j, \qquad j=1,\dots,M,$$
with the $I_j$ quantum numbers being half-integers.
From (\ref{1+a}) one can see that $i \,\log a(\theta)$ takes the value of $2\pi$ times
a half-integer number also at the positions of the zeroes of $T_1(\theta)$. Thus taking 
$i \, \log a(\theta)$ on the real axis, one can interpret this function as the counting
function of real Bethe roots, because the integer part of 
$i \, \frac{\log a(\theta')-\log a(\theta)}{2\pi}+1$ 
provides the sum of the number of those real Bethe roots and real zeroes of $T_1(\theta)$ which lie 
in the interval $[\theta,\theta']$. Using this counting function, the real zeroes of $T_1(\theta)$ 
can be considered as holes in the distribution of real Bethe roots, hence we will call them 
\emph{"type I holes"} (See figure 1.).
 The function $a(\theta)$ is important for us for the determination of real zeroes of $T_1(\theta)$ 
and for the determination of wide roots in the repulsive regime. 

We saw that the previous auxiliary function is the counting function of real roots but, since the 
ground state of the spin-1 chain is formed by 2-strings, we need to define an auxiliary function, which
can be considered in the thermodynamic limit as the counting function of 2-strings. 
The definition of such an auxiliary function is as follows \cite{1,15}:
\begin{equation} \label{b}
b(\theta)=\frac{\Phi(\theta-i\frac{\pi}{2})}{\Phi(\theta+i\frac{\pi}{2})\Phi(\theta+i\frac{3\pi}{2})}
\frac{Q(\theta+i\frac{3\pi}{2})}{Q(\theta-i\frac{3\pi}{2})} \, 
T_1\left(\theta-i\frac{\pi}{2}\right),
\end{equation}
\begin{equation} \label{B}
B(\theta)=1+b(\theta)=\frac{1}{\Phi(\theta+i\frac{\pi}{2})\Phi(\theta+i\frac{3\pi}{2})}
\frac{Q(\theta+i\frac{\pi}{2})}{Q(\theta-i\frac{3\pi}{2})} \, 
T_2(\theta).
\end{equation}
We also introduce the complex conjugate of $b(\theta)$ and $B(\theta)$ as auxiliary functions and we
denote them by ${\bar b}(\theta)$ and ${\bar B}(\theta)$ respectively.

In the thermodynamic limit the function $\frac{1}{i} \log b(\theta)$ can be considered as the
 counting function of 2-strings. The argument is as follows: from the definitions (\ref{b},\ref{B})
one can see that
 \begin{equation} \label{lnbr}
\frac{1}{i} \log b(\theta_j-i \frac{\pi}{2})=2 \pi I_j, \qquad I_j \in {\mathbb Z}+\frac12,
\qquad j=1,\dots,M,
\end{equation}
 \begin{equation} \label{lnbh}
\frac{1}{i} \log b(h_j)=2 \pi I_{h_j}, \qquad I_{h_j} \in {\mathbb Z}+\frac12,
\end{equation}
 where $h_j$ is a zero of $T_2(\theta)$.
In the large $N$ limit the 2-strings can be thought to be exact, and thus 
following from (\ref{lnbr},\ref{lnbh}), in this limit considering 
the function $\frac{1}{i} \log b(\theta)$ on the real axis one can recognize that the real part of   
$\frac{1}{i} \frac{\log b(\theta')-\log b(\theta)}{2\pi}+1$ provides the sum of the number of
real zeros of $T_2(\theta)$ lying in the interval $[\theta,\theta']$ 
and the number of 2-strings with $\theta< \mbox{Re} \, \theta_j<\theta'$.
In this way real zeroes of the transfer matrix eigenvalue $T_2(\theta)$ can be considered as
holes in the distribution of 2-strings hence we will call them simply \emph{holes}.
The holes can be depicted in the plane of effective roots, so that they lie 
on the real axis (See figure 2.).

For the closure of the NLIE we need to define a TBA type of auxiliary function as well:
\begin{equation} \label{y}
y(\theta)=\frac{T_2(\theta)}{\Phi(\theta-i\frac{3\pi}{2})\Phi(\theta+i\frac{3\pi}{2})},
\qquad Y(\theta)=1+y(\theta).
\end{equation}
An additional auxiliary function has to be defined for the determination of wide roots in the
attractive regime. Its definition is as follows:
\begin{equation} \label{h}
h(\theta)=(-1)^M \frac{\Phi(\theta-i\frac{\pi}{2}-i p \pi)}{\Phi(\theta-i\frac{\pi}{2})} \cdot
\frac{Q(\theta-i\frac{3\pi}{2})}{Q(\theta+i\frac{\pi}{2}-i p \pi)} \cdot
\frac{T_1(\theta-i\frac{\pi}{2}-i p \pi)}{T_1(\theta-i\frac{\pi}{2})}.
\end{equation}
In the context of the NLIE technique the function $h(\theta)$ turns 
out to be the most convenient to determine the positions of wide roots. It gives the quantization rule
$$ h(\theta_j-i\frac{\pi}{2})=-1 $$ for the $\theta_j$ wide roots lying on the upper half of the
 complex plane.
 
 In order to get the proper NLIE one needs to know the analytic properties of the main building
 blocks (i.e. $Q(\theta),T_1(\theta),T_2(\theta)$) of the auxiliary functions (\ref{a}-\ref{h}).
Based on numerical evidence these are as follows \cite{1,15}: \newline \newline
1. The number of zeroes of $Q(\theta)$  with imaginary parts being close to
 $\pm i \frac{\pi}{2}$ is $N$ minus $O(1)$, and it has of order 1 number of zeroes located outside of this 
region.
 \newline \newline
 2. $T_1(\theta)$ is analytic and non zero (ANZ) in the strip $\mbox{Im} \, \theta \in [-{\pi}/{2},{\pi}/{2}]$ 
apart from a finite number of zeroes located exactly on the real axis:
\begin{equation} \label{t1zh}
T_1(h^{(1)}_j)=0, \qquad j=1,\dots,N_1.
\end{equation}
3. Also $T_2(\theta)$ has only a finite number of zeroes in the strip $\mbox{Im} \, \theta \in 
[-{\pi}/{2},{\pi}/{2}]$ and they are also distributed along the real axis
\begin{equation} \label{t2zh}
T_2(h_j)=0, \qquad j=1,\dots,N_H.
\end{equation}

The knowledge of the analytic properties of these building blocks enables us to determine easily
the analytic properties of the auxiliary functions (\ref{a}-\ref{h}) which provide the starting
point for the derivation of the NLIE \cite{1,15}.

\section{The nonlinear integral equations}

In this section we present the NLIE governing the finite size scaling of the inhomogeneous 
19-vertex model with alternating inhomogeneities. Previously in \cite{1} the NLIE has been derived  
for the repulsive regime ($1<p$), and now we extend these results for the attractive regime
($0<p<1$) of the model as well. Apart from some technical subtleties, the derivation of the NLIE 
in the attractive regime is the same as in the repulsive one. Skipping the derivation of the
NLIE and referring the interested reader to references \cite{1,15}, we simply present the
final form of the NLIE:
\begin{eqnarray}
\log b(\theta) &=& C_b+i D(\theta)+i g_1(\theta)+i g_b(\theta)+(G*_{_{\Gamma}} \ln B)(\theta)-
 (G*_{_{\bar\Gamma}} \ln {\bar B})(\theta) \nonumber \\
&+& \lim_{\varepsilon \to 0^+}(K^{+\frac{\pi}{2}-\varepsilon}*\ln Y)(\theta)  \label{nlie1} \\
\log y(\theta)&=& C_y+i g_y(\theta)+(K^{+\frac{\pi}{2}}*_{_{\Gamma}} \ln B)(\theta)+
 (K^{-\frac{\pi}{2}}*_{_{\bar\Gamma}} \ln {\bar B})(\theta), \nonumber
\end{eqnarray}
where $\ln$ denotes the "fundamental" logarithm function having its branch cut on the negative real axis
and we introduced the notation for any function $f$
$$ f^{\pm \eta}(\theta)=f(\theta\pm i \eta).$$
The eqs. (\ref{nlie1}) contain three type of convolutions, one of them is the usual 
one containing integration along the real axis
$$ (f*g)(x)=\int\limits_{-\infty}^{\infty} dy \, f(x-y) g(y), $$ 
while the other two ones are defined by integrating on the complex plane along the 
integration contours $\Gamma(t)$ and $\bar\Gamma(t) \qquad (t \in \mathbb{R})$: 
$$  (f*_{_{\Gamma}} g)(x)=\int_{\Gamma} dz \, f(x-z) g(z), \qquad 
(f*_{_{\bar\Gamma}} g)(x)=\int_{\bar\Gamma} dz \, f(x-z) g(z), $$ 
where the curve $\bar\Gamma(t)$ is the complex conjugate of $\Gamma(t)$. 
The continuous non-self-intersecting contour $\Gamma(t)$ has to fulfill the following properties:  
\newline
\newline
1. $\mbox{Re}\Gamma(\pm \infty)=\pm \infty$,
\newline
\newline
2. $0\leq\mbox{Im}\Gamma(t)< \mbox{min}(1,p)\, {\pi}/{2} \qquad \forall t \in {\mathbb R} $.
\newline
\newline
The first property means that the curve $\Gamma(t)$ goes to infinity in both directions 
and the second one is necessary to avoid the poles of  the kernels $G(x)$ and $K(x)$.
Although the form of the eqs. (\ref{nlie1}) can describe correctly the finite size effects
of the spin-1 chain with any choice of the contour $\Gamma(t)$
fulfilling the previous properties but, 
since we would like to treat only few a number of roots and holes characterizing the excitations, we 
impose a third condition on $\Gamma(t)$ to be fulfilled, namely:
\newline
\newline
3. With the exception of a finite number of Bethe roots: \, $|\mbox{Im} \, 
\theta_j|<{\pi}/{2}+\mbox{Im}\Gamma(\mbox{Re} \,
\theta_j),$ \newline
or equivalently: \newline
with the exception of a finite number of effective roots: \, $|\mbox{Im} \, 
\tilde\theta_j|<\mbox{Im}\Gamma(\mbox{Re} \,
\tilde\theta_j).$
\newline
\newline
This third condition ensures that we need to treat only of order one number of roots and holes
to characterize the excitations. (See figure 2.)
 
 The kernel functions $G$ and $K$  of (\ref{nlie1}) read as
  \begin{equation} \label{GK}
 G(\theta)=\int\limits _{-\infty}^{\infty}\frac{dq}{2\pi}\,\, 
e^{iq\theta}\,\,\frac{\sinh\frac{\pi(p-1)q}{2}}{2\sinh\frac{\pi pq}{2}\cosh\frac{\pi q}{2}},\qquad 
K(\theta)=\frac{1}{2\pi\cosh(\theta)}.
\end{equation}
We also introduce the odd primitives of the kernel functions (see appendix C. for the choice of
 branch cuts of $\chi_{K}(\theta)$), 
 \begin{equation} \label{chi}
\chi(\theta)=2\pi\int\limits _{0}^{\theta}dx\,\, G(x),\qquad\chi_{K}(\theta)=2\pi\int\limits 
_{0}^{\theta}dx\,\, K(x), 
\end{equation}
that are important in writing the source terms containing information
on the excitations:
\begin{eqnarray}
g_{b}(\theta) & = & \sum_{j=1}^{N_{H}}\chi(\theta-h_{j})+
\sum_{j=1}^{N_{V}^{S}}\left(\chi(\theta-v_{j})+\chi(\theta-\bar{v}_{j})\right)
-\sum_{j=1}^{N_{S}}\left(\chi(\theta-s_{j})+\chi(\theta-\bar{s}_{j})\right) \nonumber \\
 & - & \sum_{j=1}^{M_{C}}\chi(\theta-c_{j})-\sum_{j=1}^{M_{W}}\chi_{II}(\theta-w_{j})
 -\sum_{j=1}^{M_{sc}}\chi_{II}(\theta-w_{sc}^{(j)}),\\
g_{1}(\theta) & = & \sum_{j=1}^{N_{1}}\chi_{K}(\theta-h_{j}^{(1)}),\\
g_{y}(\theta) & = & \lim_{\eta\rightarrow 0^{+}}\tilde{g}_{y}\left(\theta+i\frac{\pi}{2}-i\eta\right),
\nonumber \\
\tilde{g}_{y}(\theta) & = & 
\sum_{j=1}^{N_{H}}\chi_{K}(\theta-h_{j})
+\sum_{j=1}^{N_{V}^{S}}\left(\chi_{K}(\theta-v_{j})+\chi_{K}(\theta-\bar{v}_{j})\right)
-\sum_{j=1}^{M_{S}}\left(\chi_{K}(\theta-s_{j})+\chi_{K}(\theta-\bar{s}_{j})\right) \nonumber \\
 & - & \sum_{j=1}^{M_{C}}\chi_{K}(\theta-c_{j})
 -\sum_{j=1}^{M_{W}}\chi_{KII}(\theta-w_{j})
 -\sum_{j=1}^{M_{sc}}\chi_{KII}(\theta-w_{sc}^{(j)}), \end{eqnarray}
 where the second determination of any function: $f_{II}(\theta)$ is defined as in \cite{16}
   \begin{equation}\label{detII}
	f_{II}(\theta)= \begin{cases}
	f(\theta)+f(\theta-i\, \pi\,\mbox{sign}(\mbox{Im} \, \theta))\; & 1<p \\ 
	f(\theta) - f(\theta-i \, p \, \pi \, \mbox{sign}(\mbox{Im} \, \theta)) \; & 0<p<1. 
		   \end{cases}
\end{equation}
The objects appearing in the source terms of (\ref{nlie1}) are as follows
\newline
1. \emph{Type I holes}: \, $\{h_j^{(1)}\}, \quad j=1,\dots,N_1$
\newline
\newline
2. \emph{Holes}: \, $\{h_j\}, \quad j=1,\dots,N_H$ 
\newline
\newline
3. "\emph{Close source objects}": $\{c_j\}$ $\quad j=1,\dots,M_C$, which are "\emph{close effective roots}"
satisfying the condition: $\mbox{Im} \Gamma(\mbox{Re} \, c_j)<|\mbox{Im} \, c_j|<\mbox{min}(1,p) \, \pi.$
(See figure 2. and 3.)
\newline
\newline
4. \emph{Wide effective roots}: \, $\{w_j\}, \quad j=1,\dots,M_W$ 
\newline
\newline
5. \emph{Self-conjugated effective roots}: \, $\{w_{sc}^{(j)}\}, \quad j=1,\dots,M_{sc}$.
\newline
There are also two types of special objects in the source terms of our equations.
It is well known that special objects appear when along the integration contour 
$\Gamma(t)$ the function $B(\theta)$ crosses the cut of the logarithm function \cite{16}.

 It can be shown that special objects can always be avoided by an appropriate 
choice of the integration contour. On the other hand it can also be shown that for any eigenstate 
of the Hamiltonian the contour $\Gamma(t)$ can be chosen in such a way that the NLIE must contain 
at least one special object.
Since we want to give a formulation of the NLIE being valid for any choice of the non-self-intersecting 
integration contour $\Gamma(t)$ we need to take into account the special objects as well.

In our case the special objects $\{s_j\}$ and $\{v_j\}$ are such points of the contour $\Gamma(t)$
at which the function $\ln B(\theta)$ is exactly on the cut of the logarithm. I.e. at these points
the value of $B(\theta)$ is a negative real number
\footnote{The sets $\{\bar s_j\}$ and $\{\bar v_j\}$ denote the complex conjugates of the
sets $\{s_j\}$ and $\{v_j\}$ respectively.}.

\begin{figure}[htb]
\begin{flushleft}
\hskip 15mm
\leavevmode
\epsfxsize=160mm
\epsfbox{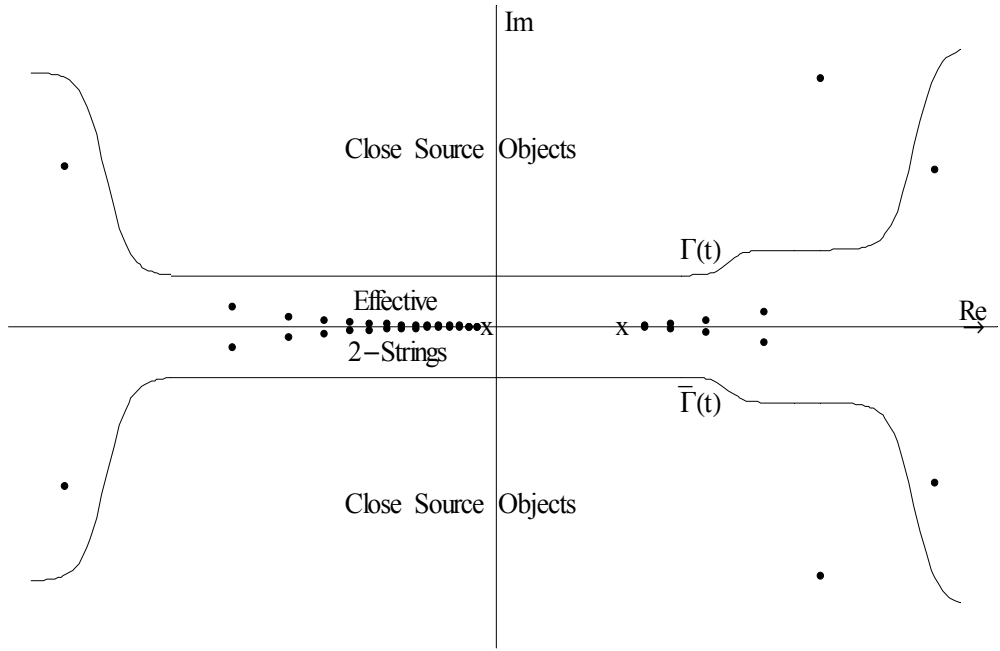}
\end{flushleft}
\caption{{\footnotesize
 Zoom of figure 2. near the real axis.
}}
\label{3}
\end{figure}
 
The difference between the two types of special objects introduced above is the sign of the derivative of
the imaginary part of the function $\log b(\theta)$ taken at the values of the special objects.
For the objects denoted by $\{s_j\}$ this derivative is negative, and for the objects $\{v_j\}$
the derivative is positive. In the rest of the paper we will call the objects $\{s_j\}$
 \emph{ordinary special objects} and we will call the objects $\{v_j\}$ \emph{virtual special objects}.
These special objects get the attribute "virtual", because as we will see later 
in many cases one can get rid of them by an appropriate choice of the integration contour $\Gamma(t).$
 
The formulation of the aforementioned conditions in the
language of the function $b(\theta)$ reads as
\begin{equation} \label{sj}
\mbox{Im} \log b(s_j)=2 \pi I_{s_j}, \quad |b(s_j)|>1, \quad 
(\mbox{Im} \log b)'(s_j)<0,
\quad j=1,\dots,N_S,
\end{equation}
\begin{equation} \label{vj}
\mbox{Im} \log b(v_j)=2 \pi I_{v_j}, \quad |b(v_j)|>1, \quad 
(\mbox{Im} \log b)'(v_j)>0,
\quad j=1,\dots,N_V^S.
\end{equation}
To provide better understanding of these special objects we demonstrate that at special
choices of the integration contour $\Gamma(t)$ the previous two types of special objects
correspond to either holes or to the upper parts of 
the effective root counterparts of the quasi 2-strings.
For the sake of simplicity hereafter we will call the close effective roots coming from
quasi 2-strings, \emph{effective 2-strings}.
When the deviations of all the quasi 2-strings are less than 
$i \, \mbox{min} (1,p) \frac{\pi}{2}$ the 
integration contour $\Gamma(t)$ can be deformed infinitesimally close to the upper parts of
the effective 2-strings and the holes. 

Let us single out the the $j_0$th 2-string with positive 2-string deviation\footnote{Only 2-strings
with positive 2-string deviations can appear in the effective plane and so in our equations.}: 
$$\theta_{j_0}^{\pm}=x_{j_0} \pm i\left(\frac{\pi}{2}+\delta_{j_0} \right), \qquad  
 x_{j_0},\delta_{j_0} \in \mathbb{R}, \qquad 0<\delta_{j_0}\ll 1.
$$
 Then the corresponding effective 2-string takes the form: $x_0^{\pm}=x_{j_0} \pm i \delta_{j_0}$. 
In the sequel let $x_0$ the position of the upper part of the effective 2-string $(x_0^+)$ or a hole.  
In this case $x_0$ fulfills the quantization condition as follows:
\begin{equation} \label{}
\log b(x_0)=2\pi I_0 \, i, \qquad I_0 \in \mathbb{Z}+1/2.
\end{equation}
Let us assume that the integration contour $\Gamma(t)$ runs infinitesimally close to $x_0$, and around this 
point behaves like $\Gamma(t)\simeq x_0+i \eta$ with $\eta$ being an infinitesimal real parameter.
Then a Taylor expansion around $x_0$ yields:
\begin{equation} \label{}
\mbox{Im} \, \log b(x_0+i \eta)=2 \pi \, I_0+\eta \, (\mbox{Re} \log b)'(x_0)+O(\eta^2),
\end{equation}
 \begin{equation} \label{}
\mbox{Re} \, \log b(x_0+i \eta)=-\eta \, (\mbox{Im} \log b)'(x_0)+O(\eta^2).
\end{equation}
Thus in the $\eta \to 0$ limit, the $\bar{x}=x_0+i \eta$ point of the contour $\Gamma(t)$ can correspond to
a special type of object (i.e $1+b(\bar{x})$ is a real negative number) in two cases:
\newline
1. $ 0<\eta$  and $(\mbox{Im} \log b)'(\bar{x})<0,$ \newline 
2. $\eta<0$ and $(\mbox{Im} \log b)'(\bar{x})>0$. 
\newline
In the first case $\bar{x}$ is an ordinary special object satisfying (\ref{sj}), and in the second case
$\bar{x}$ corresponds to a virtual special object defined by (\ref{vj}).

To summarize: from the previous simple argument it can be seen that the ordinary special objects may
appear when the integration contour runs above the positions of the upper parts of effective 2-strings or 
holes , and the virtual special objects may show up when the contour $\Gamma(t)$ runs under the upper parts 
of the effective 2-strings. 
The ordinary special objects correspond to the special objects introduced in the 6-vertex model in 
\cite{16}. 

At this point we should mention that the best choice for the integration contour when it runs above all the 
upper parts of the effective 
  2-strings and under the upper parts of the other close effective roots, but
there may be some cases (mostly in the inhomogeneous case) when due to large deviations of 
some quasi 2-strings, the effective 2-strings cannot be distinguished uniquely from other  
close effective roots, thus the differentiation between effective 2-strings and other 
close effective roots depends on our choice. 
 
 \emph{According to our choice, the integration
 contour $\Gamma(t)$ separates the effective 2-strings from those close
 effective roots which show up in our NLIE as close source objects.} 
  This definition is arbitrary and depends on the actual choice of the curve
 $\Gamma(t)$, but the NLIE (\ref{nlie1}) treats this problem consistently, and the final physical 
result, the energy is completely independent of the choice of this separating curve.
 
 The driving term bulk contribution in the equation for $\ln b(\theta)$ reads as 
 $$D(\theta)=N\arctan\frac{\sinh\theta}{\cosh\Theta}.$$
 The values of the constants of the NLIE (\ref{nlie1}) are as follows
 $$ C_b=i \, \pi \, \delta_b, \qquad \delta_b \in \{0,1 \},$$
$$ \delta_b=\frac{N}{2}+\frac{N_1}{2}-M_{sc}+S+n_- \,\, \mbox{mod} \,\, 2, \qquad 
n_-=\left[\frac{2S}{p+2}\right]-\left[\frac{S}{p+2}\right]$$
and
$$ C_y=i \, \pi \,S+ i \, \pi \, \Theta(p-1) \,(M_W+M_{sc}),$$ 
where $[...]$ stands for integer part and $\Theta(x)$ denotes the Heaviside function.
In addition to (\ref{nlie1}) we need two other equations for the determination of type I holes and 
effective wide and self-conjugated roots.
For the determination of type I holes we need to know $a(\theta)$ on the real axis:
\begin{equation} \label{lna1}
-\log a(\theta)=i \, \delta_a(\theta)+(K*_{_{\Gamma}} \ln B)(\theta)-
 (K*_{_{\bar\Gamma}} \ln {\bar B})(\theta)-C_y, \qquad 0\leq |\mbox{Im} \, \theta|< \frac{\pi}{2},
\end{equation} 
\begin{eqnarray}
\delta_a(\theta) & = & 
\sum_{j=1}^{N_{H}}\chi_{K}(\theta-h_{j})
+\sum_{j=1}^{N_{V}^{S}}\left(\chi_{K}(\theta-v_{j})+\chi_{K}(\theta-\bar{v}_{j})\right)
-\sum_{j=1}^{M_{S}}\left(\chi_{K}(\theta-s_{j})+\chi_{K}(\theta-\bar{s}_{j})\right) \nonumber \\
 & - & \sum_{j=1}^{M_{C}}\chi_{K}(\theta-c_{j})
 -\sum_{j=1}^{M_{W}}\chi_{KII}(\theta-w_{j}) -\sum_{j=1}^{M_{sc}}\chi_{KII}(\theta-w_{sc}^{(j)}).
  \end{eqnarray}
The function necessary to know for the determination of wide and self-conjugated effective roots 
is as follows, for $\mbox{min}(1,p) \, \pi<\mbox{Im} \, \theta \leq \frac{\pi(p+1)}{2}$:
\begin{eqnarray} 
\log \tilde a(\theta) &=& i \, D_{II}(\theta)+
i \, g_{1II}(\theta) +i \, g_{bII}(\theta)+(G_{II}*_{_{\Gamma}} \ln B)(\theta)-(G_{II}*_{_{\bar\Gamma}} \ln 
{\bar B})(\theta) \label{lnatilde} \nonumber \\ 
&+& ((K^{-\frac{\pi}{2}})_{II}*\ln Y)(\theta),  \label{atilde0}
\end{eqnarray}
where 
\begin{equation}\label{atilde}
	\tilde a(\theta)= \begin{cases}
	1/a(\theta+i \, \frac{\pi}{2})\; & 1<p \\ 
	1/h(\theta)  \; & 0<p<1. 
		   \end{cases}
\end{equation}
The source objects appearing in the NLIE are not arbitrary parameters, but they have to satisfy
certain quantization conditions dictated by the analytic properties of the auxiliary
functions (\ref{a}-\ref{h}). These are as follows
\begin{itemize}
\item For holes: \begin{equation}
\frac{1}{i}\,\log b(h_{j})=2\pi\, I_{h_{j}}, \qquad j=1,...,N_{H}.\label{eq:holes}\end{equation}

\item For ordinary special objects: \begin{equation}
\mbox{Im} \log b(s_j)=2 \pi I_{s_j}, \quad |b(s_j)|>1, \quad 
(\mbox{Im} \log b)'(s_j)<0,
\quad j=1,\dots,N_S.\label{eq:specials}\end{equation}

\item For virtual special objects: \begin{equation}
\mbox{Im} \log b(v_j)=2 \pi I_{v_j}, \quad |b(v_j)|>1, \quad 
(\mbox{Im} \log b)'(v_j)>1,
\quad j=1,\dots,N_V^S.\label{eq:virtspecials}\end{equation}
 
\item For close source objects (only for the upper part of the close effective pair): \begin{equation}
\frac{1}{i}\,\log b(c_{j}^{\uparrow})=2\pi\, I_{c_{j}^{\uparrow}}, \qquad 
j=1,...,M_{C}/2.\label{eq:close}\end{equation}
 
\item For wide effective roots: \begin{equation}
\frac{1}{i}\,\log \tilde{a}(w_{j}^{\uparrow})=2\pi\, I_{w_{j}^{\uparrow}}, \qquad 
j=1,...,M_{W}/2.\label{eq:wide}\end{equation}
 
\item For self-conjugated effective roots: \begin{equation}
\frac{1}{i}\,\log \tilde{a}(w_{sc}^{\uparrow(j)})=2\pi\, I_{w_{sc}^{\uparrow(j)}}, \qquad 
j=1,...,M_{sc}.\label{eq:self}\end{equation}
 
\end{itemize}
So far we have determined only the upper part of the complex pairs,
but the other parts can be determined by simple complex conjugation. 

\begin{itemize}
\item Finally for type I holes: \begin{equation}
\frac{1}{i}\,\log a (h_{j}^{(1)})=2\pi\, I_{h_{j}^{(1)}}, \qquad j=1,...,N_{1}.\label{eq:t1}\end{equation}

\end{itemize}
All the above quantum numbers $I_{\alpha_{j}}$'s are half integers.
A state is then identified by a choice of the quantum numbers $(I_{h_{j}},I_{c_{j}},...)$.
We also mention that the NLIE itself can impose constraints on the allowed values of some
of these quantum numbers.
To get the finite size effects of the spin chain one needs to solve (\ref{nlie1})
supplemented by the quantization conditions (\ref{lna1}-\ref{eq:t1}) for the two unknown functions
$b(\Gamma(t))$ and $y(x)$. The equations (\ref{lna1}-\ref{eq:t1}) also express the fact that the knowledge
of $b(\theta)$ on a single curve $\Gamma(t)$ and $y(x)$ on the real axis is enough to be able to 
express all the auxiliary functions on the whole complex plane and to determine all the roots
of the Bethe Ansatz equations (\ref{BAE}).  
 
 Having the solution of the NLIE, the corresponding energy eigenvalue of the spin chain can be expressed by it:
\begin{eqnarray} 
E &=&\frac{2 \pi^2}{\gamma} \, \bigg\{   
\sum_{j=1}^{N_{H}}{K}(h_{j})
+\sum_{j=1}^{N_{V}^{S}}\left({K}(v_{j})+{K}(\bar{v}_{j})\right)
-\sum_{j=1}^{N_{S}}\left({K}(s_{j})+{K}(\bar{s}_{j})\right)  \nonumber \\
 & - & \sum_{j=1}^{M_{C}}{K}(c_{j})
 -\sum_{j=1}^{M_{W}} K_{II}(w_{j}) -\sum_{j=1}^{M_{sc}} K_{II}(w_{sc}^{(j)}) 
 + \frac{i}{2\pi} \, \int_{\Gamma} d\theta \, K'(\theta) \, \ln B(\theta) \label{E} \\
 &-&  \frac{i}{2\pi} \, \int_{\bar\Gamma} d\theta \, K'(\theta)  \ln \bar B(\theta) 
 \bigg\}. \nonumber
 \end{eqnarray}

 \section{Counting equations}

 The counting equations are certain sum rules which give the number of holes 
 in terms of the numbers of different types of other source objects. 
 These equations play an important role in the determination of the constant
terms of the NLIE, and in the conformal analysis of the spin chain. In our spin-1 chain there are
two important counting equations: one expresses the number of holes, the other one expresses the
number of type I holes by the numbers of other source objects. 

Starting from the counting function $i \log a (\theta)$ and applying the argument of Destri and de Vega 
\cite{16} one gets the counting equation for type I holes:
\begin{equation} \label{cet1}
N_1-2N_R^S=S+M_1+M_{sc}-M_R-2n_-,
\end{equation}
where $M_1$ denotes the sum of effective roots with the property 
$\frac{\pi}{2}<|\mbox{Im} \, \tilde\theta_j|<\frac{\pi(p+1)}{2}$, 
$M_R$ stands for the number of real Bethe roots, and $N_R^S$ means the number of the \emph{real special 
objects}.
Real special objects can be either real Bethe roots or type I holes. They are called specials because
at the positions of real specials the counting function of real Bethe roots is no more monotonically  
increasing: i.e. $i \, \frac{d}{dx} \log a (x)<0$.
 
 There is also a parity constraint for the number of type I holes. Namely \emph{$N_1$ must be even}.
 The proof of this statement is related to the derivation of the second counting equation. 
  
    From the definition (\ref{b})
 of $b(\theta)$ it can be shown that the difference $\frac{1}{i} \, (\log b(\infty)-\log b(-\infty))$ 
must be equal to $2 \pi$ times an integer. Calculating this difference from the equations (\ref{nlie1}), 
one gets that $\frac{1}{i} (\log b(\infty)-\log b(-\infty))=N_1 \pi
+2 \pi \times (\mbox{integer})+\frac{2 \pi \left(  
N_H+2N_V^S-2N_S-M_C-2S-2\Theta(p-1)(M_W+M_{sc})+2N_{+}\right)}{p}.$
Matching the two conditions for any values of $p$ is possible only if $N_1$ is even and
the expression divided by $p$ in the last formula is equal to zero. This leads to the
second counting equation:
\begin{equation} \label{cet2}
N_H+2N_V^S-2N_S =M_C+2S+2\Theta(p-1)(M_W+M_{sc})-2N_{+},
\end{equation}
where $N_{+}=\left[ \frac{3S}{p+2}\right]-\left[ 
\frac{S}{p+2}\right]$.
In case of the absence of virtual special objects the form of this counting equation is very similar
 to the one of the six-vertex model \cite{16}.

\section{Excitations in the large $N$ limit}

In this section using our NLIE (\ref{nlie1}) we discuss qualitatively the typical Bethe root 
configurations of the spin chain (homogeneous case) in the $N \to \infty$ limit.  
 In this limit the 2-string deviations are small hence the integration contour $\Gamma(t)$
can go above the positions of all the effective 2-strings. Moreover the counting function
$\frac{1}{i} \log b(\theta)$ is dominated by the bulk source function $D(\theta)$
as far as $-\ln N \lesssim \mbox{Re} \, \theta \lesssim \ln N$. In this intermediate regime
$\ln B(\theta)$ is exponentially small in $N$, hence the integral terms containing $\ln B(\theta)$ 
can be dropped. The integrals containing $\ln Y(\theta)$ cannot be dropped, but their contribution
is of order one, so their contribution is only next to leading order in the large $N$ limit. Their contribution 
is relevant for the determination of 2-string deviations, but irrelevant for the discussion of the
qualitative root positions of the large $N$ limit. 
Hereafter we need to discuss separately the attractive and repulsive 
regimes of the model, because the typical root configurations other than quasi 2-strings are completely
different.   

\subsection{Repulsive regime $1<p$}

According to (\ref{eq:close}) the positions of close source objects are to be determined from the 
function
$\frac{1}{i} \log b(\theta)$. The quantization condition (\ref{eq:close}) also tells us that the 
imaginary part of
$\frac{1}{i} \log b(\theta)$ at the positions of close source objects must be equal to zero. However
the driving term $D(\theta)$ has large imaginary part in the large $N$ limit, which can be compensated 
only if the relative positions of close source objects tend to certain singularities of source function 
$g_b(\theta)$. The singularities of $g_b(\theta)$ comes from the function $\chi(\theta)$ which has 
poles at $\pm i \pi$. 

Hence the cancellation of the (of order $N$) contribution to the imaginary part of 
$\frac{1}{i} \log b(\theta)$ dictates that the relative positions of close source objects must tend to 
$i \pi$. Consequently, in the large $N$ limit, close source objects form pairs with difference
$\pi$ in thier imaginary parts:
$$ c^+=c_0+i \left( \frac{\pi}{2}+\mu \right), \qquad c^-=c_0-i \left( \frac{\pi}{2}-\mu \right),
	 \qquad c_0,\mu \in \mathbb{R}, \quad 0<\mu<\frac{\pi}{2}. $$
On the other hand it is known that close source objects appear in complex conjugated pairs, thus 
they must fall either into 2-string configurations on the effective plane ($\mu=0$):
$c^{\uparrow \downarrow}=c_0\pm i \frac{\pi}{2}$, or into \emph{"effective quartets"} with:
$c^{\pm \uparrow \downarrow}=c_0\pm i \left(\frac{\pi}{2} \mp \mu \right)$,
just like in the spin-1/2 XXZ chain \cite{26}.
	 	 
 However our NLIE does not give any constraint on the imaginary parts of the wide roots, because there
 is no bulk driving term in (\ref{atilde0}). This is because $D_{II}(\theta)=0$ in the repulsive regime.
Transforming back the effective roots to the language of original Bethe roots, one gets that in the 
large $N$ limit of the spin chain the possible root configurations are as follows
\newline
\newline
1. \emph{Real roots:} $\mbox{Im} \, \theta_j=0,$
\newline
\newline
2. \emph{Inner roots:} $0<|\mbox{Im} \, \theta_j|<\frac{\pi}{2},$
\newline
\newline
3. \emph{Quasi 2-strings:} $\mbox{Im} \, \theta_j \simeq \pm \frac{\pi}{2},$
\newline
\newline
4. \emph{Close roots with:} $\mbox{Im} \, \theta_j \simeq \pm \pi,$
\newline
\newline
5. \emph{Close quartets with:}  $\theta^{\pm \uparrow \downarrow}=\theta_0 \pm i \left( \pi \mp \mu 
\right), \qquad 0<\mu<\frac{\pi}{2}, \quad \theta_0 \in \mathbb{R},$
\newline
\newline
6. \emph{Wide roots} with arbitrary imaginary parts,
\newline
\newline
7. \emph{Self-conjugated roots} with: $\mbox{Im} \, \theta_j= \frac{\pi(p+2)}{2}.$

This means that in the large $N$ limit,
contrary to the prediction of the string hypothesis \cite{8,9,10}, in the spectrum of Bethe roots
 at best only 1-, 2-, and 3-strings and anti-strings accompanied by 
 quartets, wide- and inner-roots
 without constrained imaginary parts can be seen. 
 In this regime there are no constraints on the positions of wide- and inner-roots, because of the absence 
of driving bulk source terms in (\ref{lna1}) and (\ref{atilde0}).

\subsection{Attractive regime $0<p<1$}

Contrary to the repulsive regime in the attractive case the bulk driving term $D_{II}(\theta)$ 
of $\log \tilde a (\theta)$ is no longer zero imposing further constraint on the imaginary parts
of wide roots as well. Analyzing the NLIE and following the argumentation of \cite{16}, the effective 
roots fall into the configurations as follows:
\newline
\newline
1. \emph{Arrays of the first kind} are effective root configurations containing close source objects as 
well. These type of arrays consist of effective roots of the form:
$$ \tilde{\theta_k}^{(1)\pm}=\theta \pm i \left(\mu-k p \pi \right), \quad 
\tilde{\theta_k}^{(2)\pm}=\theta \pm i \left( \pi-\mu-(k-1)p \,\pi \right),
\quad k=0,\dots,\left[\frac{1}{2p} \right],$$
with $\theta$ and $0<\mu$ being real parameters. At certain special values of $\mu$ these arrays
degenerate. 
There are two degenerate cases: \emph{odd degenerate arrays}, which contain a self-conjugated effective root at 
$$ \tilde\theta_{sc}=\theta+i \frac{\pi(p+1)}{2}, \qquad \quad \theta \in \mathbb{R}$$ and
accompanying complex pairs at
$$ \tilde\theta_k=\theta \pm i \frac{\pi (1-(2k+1)p)}{2}, \quad k=0,\dots,\left[\frac{1}{2p} \right],$$
and \emph{even degenerate ones}, which contain complex-pairs of effective roots at the positions
$$ \tilde\theta_k=\theta \pm i \frac{\pi (1-2 k p)}{2}, \quad k=0,\dots,\left[\frac{1}{2p} \right].$$
These degenerate arrays contain exactly one pair of close source objects.
\newline
\newline
2. \emph{Arrays of the second kind} contain only wide- and self-conjugated effective roots.
The \emph{odd ones} contain a self-conjugated effective root 
$$ \tilde\theta_{sc}=\theta+i \frac{\pi(p+1)}{2}, \qquad \quad \theta \in \mathbb{R}$$ and wide effective-pairs 
at
$$ \tilde\theta_k=\theta \pm i \frac{\pi (1-(2k+1)p)}{2}, \quad k=0,\dots,s, \quad
0\leq s \leq \left[\frac{1}{2p} \right]-1,$$
while the \emph{even ones} contain only wide effective-pairs
$$ \tilde\theta_k=\theta \pm i \frac{\pi (1-2 k p)}{2}, \quad k=0,\dots,s, \quad
0\leq s \leq \left[\frac{1}{2p} \right]-1.$$
The deviations of the imaginary parts of the previous effective root configurations from the 
formulae listed above are exponentially small in $N$. 


Transforming back the effective roots into original Bethe roots we get the following possible
Bethe root configurations in the attractive regime of the model:
\newline
\newline
1. \emph{Real roots:} $\mbox{Im} \, \theta_j=0,$
\newline
\newline
2. \emph{Inner roots:} $0<|\mbox{Im} \, \theta_j|<\frac{\pi}{2},$
\newline
\newline
3. \emph{Quasi 2-strings:} $\mbox{Im} \, \theta_j \simeq \pm \frac{\pi}{2},$
\newline
\newline
4. \emph{Arrays of the first kind} with Bethe roots at positions:

$$ \left.
\begin{array}{ll}  
{\theta_k}^{(1)\pm}=\theta \pm i \left( \frac{\pi}{2}+\mu-k p \pi \right)  \\ 
{\theta_k}^{(2)\pm}=\theta \pm i \left( \frac{3\pi}{2}-\mu-(k-1)p \,\pi \right) 
\end{array} \right\} 
\quad  k=0,\dots,\left[\frac{1}{2p} \right].$$
5. \emph{Odd degenerate arrays of the first kind}: 
$$ \theta_{sc}=\theta+i \frac{\pi(p+2)}{2}, \quad
 \theta_k=\theta \pm i \frac{\pi (2-(2k+1)p)}{2}, \quad k=0,\dots,\left[\frac{1}{2p} \right],$$
6. \emph{Even degenerate arrays of the first kind:}
$$  \theta_k=\theta \pm i \frac{\pi (2-2kp)}{2}, \quad k=0,\dots,\left[\frac{1}{2p} \right],$$
7. \emph{Odd arrays of the second kind}:
$$ \theta_{sc}=\theta+i \frac{\pi(p+2)}{2}, \quad \theta_k=\theta \pm i \frac{\pi (2-(2k+1)p)}{2}, \quad 
k=0,\dots,s, \quad 0\leq s \leq \left[\frac{1}{2p} \right]-1,$$
8. \emph{Even arrays of the second kind:}
$$ \theta_k=\theta \pm i \frac{\pi (2-2 k p)}{2}, \quad k=0,\dots,s, \quad 0\leq s \leq \left[\frac{1}{2p} 
\right]-1.$$

Finally we mention that the NLIE does not contain the inner roots as source objects, thus
it does not impose any constraint on the positions of this type of Bethe roots. 
Knowing the solution of the NLIE their positions
can be determined from the counting function $i \log a(\theta)$ of (\ref{lna1}) by imposing the 
quantization condition
\begin{equation} \label{}
i \log a(\theta_j^{(I)})=2 \pi I_{j}^{(I)}, \qquad I_{j}^{(I)} \in \mathbb{Z}+\frac12,
\end{equation} 
where the positions of inner roots are denoted by $\theta_j^{(I)}$.


\section{2-string deviations}

In this section we discuss the 2-string deviations for the ground state and for some simple
excited states of the repulsive regime. The method of the calculation of 2-string deviations
relies on the NLIE technique, and  can be extended easily in principle for all excited states 
of the model. 

Previously the leading large $N$ corrections of the 2-string deviations of the ground state of the spin-1 
XXZ chain were calculated analytically in \cite{11,12}. Now we extend these calculations for excited states as 
well. Let us introduce the counting function
\begin{equation} \label{Zb}
Z_b(\theta)=\frac{1}{i \, N} \log b(\theta),
\end{equation}
and write the quasi 2-strings as $\theta_j=x_j\pm i (\frac{\pi}{2}+\delta_j),$ where
$x_j$s are the string centers and $\delta_j$s are the 2-string deviations in imaginary direction. 
Then following from the definition (\ref{b},\ref{B}) of $b(\theta)$ the quantization conditions hold as 
follows:
\begin{equation} \label{Zbq}
Z_b(x_j+i \delta_j)=\frac{2 \pi I_{j}}{N}.
\end{equation}
Let us introduce the functions
\begin{equation} \label{calK}
{\cal K}(\theta)=\lim_{\varepsilon \to 0^+}(K^{+\frac{\pi}{2}-\varepsilon}*\ln Y)(\theta),
\end{equation}
\begin{equation} \label{calKb}
{\cal K}_b(\theta)=\frac1i(G*_{_{\Gamma}} \ln B)(\theta)-\frac1i
 (G*_{_{\bar\Gamma}} \ln {\bar B})(\theta).
\end{equation}
Then using (\ref{nlie1}-\ref{detII}) $Z_b(\theta)$ can be written as 
\begin{equation} \label{zbz0}
Z_b(\theta)=Z_0(\theta)+\frac{1}{i \,N} \mbox{Re} \, {\cal K}(\theta).
\end{equation}
where
\begin{equation} \label{z0}
Z_0(\theta)= 
 \arctan\sinh \theta+ \frac{\pi \delta_b+g_1(\theta)+g_b(\theta)
 +{\cal K}_b(\theta)+\mbox{Im} \, {\cal K}(\theta)}{N}.
\end{equation}
It is important to remark that $Z_0(\theta)$ is always real on the real axis. Analyzing the formulae
(\ref{zbz0}) and (\ref{z0}), one can recognize that in the large $N$ limit the driving contribution
for $Z_b(\theta)$ is in $Z_0(\theta)$, which is real along the real axis and the imaginary contribution
coming from the term $\frac{1}{i \,N} \mbox{Re} \, {\cal K}(\theta)$ is only of order $1/N$ giving 
only small correction to $Z_0(\theta)$. 
From now on we assume that the 2-string deviations are small ($|\delta_j|\ll 1$) thus we can Taylor
expand (\ref{Zbq}) around the 2-string centers:
\begin{equation} \label{ZbqT}
\frac{2\pi I_j}{N}=Z_0(x_j)+i \, \delta_j \, Z_0'(x_j)+\frac{1}{i \,N} \mbox{Re} \, {\cal K}(x_j)+
\frac{\delta_j}{N} (\mbox{Re} \, {\cal K})'(x_j)+O(\delta^2),
\end{equation}
Assuming that $\delta_j$s are of order $1/N$ and requiring the equality up to this order one gets:
\begin{equation} \label{q1}
\frac{2\pi I_j}{N}=Z_0(x_j),
\end{equation}
\begin{equation} \label{q2}
\delta_j=\frac{1}{N} \frac{\mbox{Re} \, {\cal K}(x_j)}{Z_0'(x_j)}.
\end{equation}
The last formula (\ref{q2}) provides the analytical formula for the 2-string deviations. 
As it can be seen from (\ref{q1}) the function $Z_0(\theta)$ yields the quantization rules
for the 2-string centers, thus it can be considered as counting function of 2-string centers.
It follows that the function $Z_0'(\theta)$ is the density function of 2-string centers, and so
the 2-string deviations are inversely proportional to $N$ times the 2-string density just as it was 
experienced in case of the ground state in \cite{11}, hence 
our approximation is really valid as far as the density is not too small (in the middle of the
distribution). 

To get concrete formulae for the 2-string deviations of certain excited states one has to 
solve (\ref{nlie1}) in the large $N$ limit. Then the contribution
of the function ${\cal K}_b(\theta)$ is negligible since it  tends to zero exponentially in $N$.
On the other hand in the large $N$ limit the contribution of ${\cal K}(\theta)$ is of order 1 providing  
2-string deviations. 
\newpage
Let us discuss a few examples:
\newline
{\bf Ground state} 
\newline
As it is well known the ground state is formed by only 2-strings and using a $\Gamma(t)$ contour 
with appropriately large imaginary part there are no source terms present in the NLIE (\ref{nlie1}).  
Analyzing the NLIE (\ref{nlie1}), in the large $N$ limit it turns out that  $y(\theta)=1$ 
and ${\cal K}(\theta)=\frac12 \ln 2$. Substituting these results into the formulae (\ref{z0}),
(\ref{q1}) and (\ref{q2}) one gets the well known formula of \cite{11}:
\begin{equation} \label{gs2str}
\delta_j=\frac{\ln 2}{2N}\, \cosh x_j.
\end{equation}
From this formula one can see that the 2-string deviations can be considered to be small as long as 
$|x_j|\lesssim \ln N.$  
\newline \newline
{\bf 2-hole states in the repulsive regime}
\newline
From (\ref{cet2}) one can see that holes can appear only in pairs thus the excitations
with smallest energies contain 2-holes. According to (\ref{cet1}) and (\ref{cet2}) the two holes can be  
accompanied by a pair of close source objects or by a single self-conjugated root and
for these configurations the number of type I holes is zero $(N_1=0).$
Let $h_1$ and $h_2$ the positions of the holes and assume that $h_1<h_2$. Then from the large $N$ analysis 
of the NLIE one gets for all the possible configurations containing 2-holes that
$$y(\theta)=\tanh\left(\frac{\theta-h_1}{2} \right)\tanh\left(\frac{\theta-h_2}{2} \right).$$
Then the evaluation of ${\cal K}(\theta)$ \cite{6} yields
\begin{equation} \label{K2}
{\cal K}(\theta)=i \, \pi \, \left\{ 
Q_1\left(\theta-\frac{h_1+h_2}{2}\right)-Q_2(\theta-h_1)-Q_2(\theta-h_2 ) \right\},
\end{equation}
where 
\begin{equation}
Q_1(\theta)=-\frac{1}{2 \pi} \arctan \sinh (\theta)-\frac{i}{2 \pi}
\ln \cosh(\theta),
\end{equation}
\begin{equation}
Q_2(\theta)=-\frac{\chi_2(\theta)}{2 \pi}-\frac{i}{2 \pi}\ln \cosh\left(\frac{\theta}{2}\right),
\end{equation}
and $\chi_2(\theta)$ denotes $\chi(\theta)$ of (\ref{chi}) calculated at $p=2$.
Knowing ${\cal K}(\theta)$ and the actual root configuration accompanying the two holes
it is easy to calculate $Z_0(\theta)$ in the large $N$ limit. Here we do not implement this calculation 
but rather we concentrate on the calculation of the numerator of (\ref{q2}), because the sign of the 
numerator yields the sign of the 2-string deviations. Taking the real part of (\ref{K2}) along the real 
axis one gets
\begin{equation} \label{rpK}
\mbox{Re}{\cal K}(\theta)=\frac12 \ln \cosh\left( \theta-\frac{h_1+h_2}{2}\right)-
\frac12 \ln \cosh\left( \frac{\theta-h_1}{2} \right)-\frac12 \ln \cosh\left( 
\frac{\theta-h_2}{2}\right).
\end{equation}
Analyzing (\ref{rpK}) together with (\ref{q2}) it turns out that the 2-string deviations are
negative between the  positions of the two holes 
(i.e. $\delta_j<0,$  if  $h_1<x_j<h_2$), and the 2-string deviations are positive outside of this region
(i.e. $\delta_j>0,$ if $h_2<x_j$  or $x_j<h_1$).
 
Thus we have shown for all the 2-hole states that the 2-string deviations change sign at the positions 
of holes. Relying on earlier numerical studies of ref. \cite{22}
we assume that this statement persists for multi hole states as well. 
To summarize, we can say that the 2-string deviations can be positive and negative as well, and the 
2-string deviations change their signs at the positions of the holes. For a numerical example 
see figure 4.

However we should remark that our derivation and so the above statement concerning the relation between 
the positions of holes and the signs of 2-string deviations is valid only for the middle of the root 
distribution, where the roots are densely distributed (i.e $|\mbox{Im} \, \theta| \lesssim \ln N$).
Outside of this region nothing definite can be said. For example in the presence of ordinary special roots, 
which are 
typically located at the rarely distributed edges of the root distribution, there are holes at the 
positions of which 2-string deviations do not change sign. This can be naively explained by (\ref{q2})
saying that at the positions of ordinary special roots ($x_j^S$) the density of 2-strings is negative (i.e. 
$Z_0'(x_j^S)<0$), thus not only the numerator, but also the denominator of (\ref{q2}) changes sign at the 
positions of certain holes. 

Just to mention an example for this case:
in the inhomogeneous case there exist a numerical example where $N_S=1, \quad N_H=2$ and the two holes 
induced by the single ordinary special object do not change the sign of the 2-string deviations. (See figure 5. 
in 
appendix C. of ref. \cite{6}.) 
\begin{figure}[htb]
\begin{flushleft}
\hskip 15mm
\leavevmode
\epsfxsize=160mm
\epsfbox{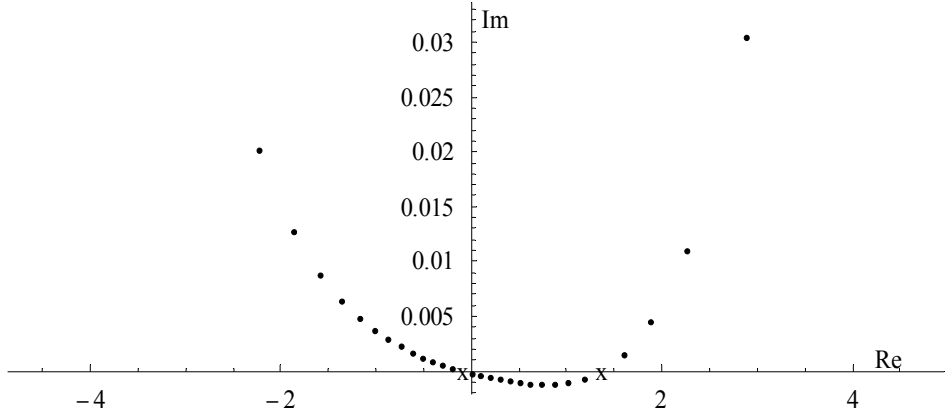}
\end{flushleft}
\caption{{\footnotesize
Deviations of the upper parts of 2-strings in a pure 2-hole state $(S=1)$ at $p=2,$ $N=64$ with
hole quantum numbers $I_{h_1}=-1/2$ and $I_{h_2}=12+1/2.$
}}
\label{4}
\end{figure}

\section{Conformal limit}

In this section for $N$ even we calculate analytically the conformal spectrum of the spin chain.
It is well known that the energy of the lowest lying excitations form a conformal spectrum in the
large $N$ limit, i.e. the energy levels behave as 
\begin{equation} \label{}
E\simeq \frac{2\pi^2}{\gamma} \left( -\frac{c}{12}+\Delta^++\Delta^- \right)\frac{1}{N},
\end{equation}
where $c$ is the central charge and $\Delta^\pm$ are the left and right conformal dimensions
of the underlying conformal field theory.
 In this thermodynamic limit the energy levels can be evaluated without solving the NLIE explicitly
 \cite{12,16,21}. For the analytical calculation of the conformal spectrum we follow the approach 
 described in detail in \cite{16}.
The positions of the sources for $N \rightarrow \infty$ can remain finite
(central objects), or they can move towards the two infinities as
$\pm\ln N$ (left\emph{/}right movers). We
introduce the finite parts $\tilde\theta_{j}^{\pm,0}$ of their positions
$\tilde\theta_{j}$ by subtracting the divergent contribution: \[
\{\tilde\theta_{j}\}\rightarrow\left\{ \tilde\theta_{j}^{\pm}\pm\ln N,\,\tilde\theta_{j}^{0}\right\} .\]
 We denote the number of right/left moving and central objects by
$N_{H}^{\pm,0},N_{S}^{\pm,0},M_{C}^{\pm,0},\dots$ etc. For later
convenience we introduce the right/left moving and central spin given
by \begin{equation}
S^{\pm,0}=\frac{1}{2}\left(N_{H}^{\pm,0}+2N_V^{S,\pm,0}-2N_{S}^{\pm,0}-M_{C}^{\pm,0}-
2 \, \Theta(p-1)\, (M_{W}^{\pm,0}+M_{sc}^{\pm,0})\right).\end{equation}
According to (\ref{cet2}) they satisfy the sum rule:
$$ S^++S^-+S^0=S-N_+.$$
 In the thermodynamic limit the NLIE splits into three separate equations corresponding
to the three asymptotic regions. This is why for all the auxiliary
functions of the NLIE (\ref{a}-\ref{h}) we define the so called
\emph{kink functions} as \begin{equation}
F_{\pm}(\theta)=\lim_{N \rightarrow \infty}F\left(\theta\pm\ln N \right).\qquad F\in\{\log b,\,\log 
y,\log\tilde{a},\dots\}.\end{equation}
 In the large $N$ limit these kink functions satisfy the so called \emph{kink
equations} and the energy and momentum can be expressed by them. Performing
the above \emph{kink limit} on our NLIE the kink equations take the
form \begin{eqnarray}
\log b_{\pm}(\theta) & = & C_b + i\, C_{b\pm} \pm i\, e^{\pm\theta}+i\, g_{1\pm}^{(0)}(\theta)
+i\, g_{b\pm}^{(0)}(\theta)\nonumber \\
 &+& (G*_{_{\Gamma_\pm}}\ln B_{\pm})(\theta)-(G*_{_{\bar{\Gamma}_\pm}}\ln\bar{B}_{\pm})(\theta)
+(K^{-\frac{\pi}{2}+\epsilon}*\ln Y_{\pm})(\theta), \label{kinkb} \end{eqnarray}
\begin{equation}
\log y_{\pm}\left(\theta-i\frac{\pi}{2}\right)=C_y+\tilde{C}_{y\pm}+i \, \tilde{g}_{y\pm}^{(0)}(\theta)
+(K*_{_{\Gamma_\pm}} \ln B_{\pm})(\theta)-(K*_{_{\bar{\Gamma}_\pm}}  
\ln\bar{B}_{\pm})(\theta),\end{equation}
\begin{equation}
\log y_{\pm}(\theta)=\lim_{\eta\rightarrow{\frac{\pi}{2}}^{-}}\log 
y_{\pm}\left(\theta-i\frac{\pi}{2}+i\eta\right).\label{lny1pmdef}\end{equation}
\begin{equation}
-\log a_{\pm}\left(\theta \right)=C_y+\tilde{C}_{y\pm}+i \, \delta_{a\pm}^{(0)}(\theta)
+(K*_{_{\Gamma_\pm}} \ln B_{\pm})(\theta)-(K*_{_{\bar{\Gamma}_\pm}}  
\ln\bar{B}_{\pm})(\theta),\end{equation}
\begin{equation} \label{kinkatilde}
\ln{\tilde{a}}_{\pm}(\theta)=i 
C_{\tilde{a}}^{\pm}+i\,\tilde{g}_{\tilde{a}\pm}^{(0)}(\theta)+(G_{II}*_{_{\Gamma_\pm}} \ln 
B_{\pm})(\theta)-(G_{II}
*_{_{\bar{\Gamma}_\pm}} \ln\bar{B}_{\pm})(\theta),\end{equation}
 where the functional forms of the source functions of the kink functions (\ref{kinkb}-\ref{kinkatilde}) 
are the same as those of the original equations (\ref{nlie1}-\ref{atilde}) with the
 difference that they contain only the finite part of the left- and right-moving holes and
 effective roots respectively. \footnote{Left moving kink functions contain only the left moving 
 (indexed by -) objects, and 
the right moving kink functions contain only the right moving (indexed by +) objects in their
source terms.} (See for example \cite{6}). The bulky expressions of the constant terms of the equations 
(\ref{kinkb}-\ref{kinkatilde}) are listed in appendix B.
The finite part of the right and left
moving objects can be obtained from the kink functions by imposing
quantization conditions very similarly to (\ref{eq:holes},...,\ref{eq:t1}).

After some manipulations \cite{6,12,16,21} it turns out 
that in the thermodynamic limit the energy and momentum can be expressed by a
sum of dilogarithm functions with the $\theta\rightarrow\pm\infty$
limiting values of the kink functions in their argument. One group of these
 limiting values agrees with the limiting values of the original auxiliary functions at the
infinities, namely: 
\begin{equation}
b_{\pm}(\pm\infty)=e^{\pm 3 i\gamma S}\,2\,\cos(\gamma S), \quad B_{\pm}(\pm\infty)=e^{\pm2i \gamma 
S}\,\frac{\sin\left(3 \gamma S \right)}{\sin\left(\gamma S \right)}
,\label{ap1}\end{equation}
\begin{equation}
y_{\pm}(\pm\infty)=\frac{\sin\left(3\gamma S\right)}{\sin\left(\gamma S\right)},\qquad 
Y_{\pm}(\pm\infty)=4\,{\cos(\gamma S)}^{2}>0.\label{ap3}\end{equation}
The other group of limiting values of the kink functions can be determined from the kink
equations (\ref{kinkb}-\ref{kinkatilde}). They read as follows
\begin{equation}
b_{\pm}(\mp\infty)=\bar{b}_{\pm}(\mp\infty)=0,\end{equation}
\begin{equation}
y_{\pm}(\mp\infty)=(-1)^{\delta_{y}^\pm}, \qquad \delta_{y}^\pm=N_+ +2S^{\pm} \, \, \mbox{mod} \, 2.
\end{equation}
 Using the dilogarithm sum rule of appendix A, and putting everything
together one gets that the central charge is $3/2$ and
the conformal weights take the form 
\begin{equation} \label{Dpm}
\Delta^{\pm}=\frac{1}{16}\,\delta_{y}^\pm+\frac{1}{2}\left(\frac{Q_{\pm}}{R}+\frac{S}{2}R\right)^{2}+
\tilde{N}_{\pm}+J_{\pm},\end{equation}
 where 
 \begin{equation}
 Q_\pm=S^\pm-\frac{S-N_+}{2},
 \end{equation}
 \begin{equation}
\tilde{N}_{\pm}=\frac{\hat{N}_{+}-\delta_{y}^\pm}{8}+S S^\pm-\frac32 (S^\pm+N_+)S^\pm
.\label{Npm}\end{equation}
\begin{eqnarray}
J_{\pm} & = & \mp I_{h^{\pm}}\mp 2I_{v^{\pm}}\pm2I_{s^{\pm}}\pm2I_{c^{\pm\uparrow}}\pm 
2I_{w^{\pm\uparrow}}\pm I_{w_{sc}^{\pm\uparrow}}\mp I_{h^{(1)\pm}} \label{Jpm} \\
 &+& S^{\pm} (M_{sc}^\pm-M_{sc}+\frac{N_1}{2}-N_1^\pm+\frac{N}{2}\pm \delta_b)+
 N_{II}^\pm+\frac12 (M_{sc}^{\pm})^2,
 \nonumber \end{eqnarray}
and 
$$I_{h^{\pm}}=\sum\limits_{j=1}^{N_H^\pm} I_{h_j^\pm}, \quad
 I_{s^{\pm}}=\sum\limits_{j=1}^{N_S^\pm} I_{s_j^\pm}, \quad
 I_{v^{\pm}}=\sum\limits_{j=1}^{N_V^{S\pm}} I_{v_j^\pm}, \quad
I_{c^{\pm\uparrow}}=\sum\limits_{j=1}^{M_C^{\pm\uparrow}}  I_{c_j^{\pm\uparrow}},\dots 
\mbox{etc.}$$
furthermore
$\hat N_+=N_+ \, \mbox{mod} \, 2,$  
and $N_{II}^\pm$ are integers depending on the relative positions 
of complex effective roots and  they must be determined case by case. 
In (\ref{Dpm}) the parameter $R$ is the compactification radius defined by
\begin{equation} \label{R}
R=\sqrt{\frac{p}{p+2}}.
\end{equation}
In order that the formulae (\ref{Dpm}-\ref{R}) describe the conformal weights of a $c=3/2$ conformal field 
theory, the difference $\Delta^+-\Delta^-$ must be independent of $R$. This requirement suffices, 
because in the conformal limit the momentum eigenvalues of the spin chain behaves as 
$P \sim (\Delta^+-\Delta^-)/N$, on the other hand the operator $e^{iP}$ is the
 operator which translates a state by one lattice site. Thus due to the periodic boundary conditions
 after $N$ unit translations one gets back the original state, i.e. $e^{iPN}=1$.
 It follows that the eigenvalues of the momentum operator look like
   $P=\frac{2 \pi}{N} \times \mbox{integer}$ hence the difference $\Delta^+-\Delta^-$ is independent of the 
compactification radius $R$. 
The fact that $\Delta^+-\Delta^-$ is independent of the compactification radius implies some
constraints on the possible root configurations in the thermodynamic limit.
The independence of $R$ can be implemented in two ways:
\newline \newline
1. $Q_+=Q_-$ in this case $S^0$ is zero. 
\newline
2. $Q_+=-Q_-$ in this case $S^+=S^-$. 
\newline \newline
To summarize: in the thermodynamic limit for states describing the lowest excitations
\footnote{Roughly speaking "lowest excitations" means 
that these excitations give only of order $1/N$ correction to the energy of the ground state.
In the repulsive regime this means that $N_H^0=0$, because 
central holes would give an of order one contribution to the energy. In case of the attractive regime
also the number of the central wide effective roots is zero when we speak about "lowest excitations" in
the large $N$ limit.}
only such root configurations appear which carry either zero 
central spin ($S^0=0$) or their left and right moving spin is equal $(S^+=S^-)$.  

Analyzing the formulae (\ref{Dpm}-\ref{R}) one can see that the spin of the state can be identified
with the winding number of the Gaussian part of the $c=3/2$
CFT, and the parameters $\delta_y^{\pm}$ distinguish the 
Neveu-Schwartz and Ramond sectors. Depending on the state under consideration
 the sum $\tilde N_{\pm}+J_{\pm}$ can be either integer or half integer, but in the Ramond sector they
 are always integers. Furthermore the $S^0=0$ and $S^+=S^-$ conditions guarantee that 
$\delta_y^{+}=\delta_y^{-}$ ensuring the correct appearance of the $1/16$ terms in the conformal weights. 
Moreover further analysis of (\ref{Dpm}) shows that there is a relation between $Q_\pm$  and
the winding number $S$, namely in the NS sector ($\delta_y^{\pm}=0$):
\begin{equation}
Q_{\pm}\in\mathbb{Z}\qquad\mbox{if}\qquad S\in2\mathbb{Z},\label{f1ns}\end{equation}
\begin{equation}
Q_{\pm}\in\mathbb{Z}+\frac{1}{2}\qquad\mbox{if}\qquad S\in2\mathbb{Z}+1,\label{f2ns}\end{equation}
 while in the Ramond sector ($\delta_y^{\pm}=1$):
 \begin{equation}
Q_{\pm}\in\mathbb{Z}+\frac{1}{2}\qquad\mbox{if}\qquad S\in2\mathbb{Z},\label{f1r}\end{equation}
\begin{equation}
Q_{\pm}\in\mathbb{Z}\qquad\mbox{if}\qquad S\in2\mathbb{Z}+1.\label{f2r}\end{equation}
 Putting together the results of the previous analysis of the conformal weights (\ref{Dpm}-\ref{R}),
one can recognize that they can be interpreted within the framework of $c=3/2$ CFT as conformal weights
appearing in the modular invariant partition function of Di Francesco et al. \cite{24}:   
\begin{eqnarray*}
Z(R) & = & \frac{1}{|\eta|^{2}}\left\{ 
(\chi_{0}\bar{\chi}_{1/2}+\chi_{1/2}\bar{\chi}_{0})\sum_{n\in\mathbb{Z}+\frac{1}{2},\, 
m\in2\mathbb{Z}+1}\right.\\
 & + & \left.(|\chi_{0}|^{2}+|\chi_{1/2}|^{2})\sum_{n\in\mathbb{Z},\, 
m\in2\mathbb{Z}}+|\chi_{1/16}|^{2}\sum_{2n-m\in2\mathbb{Z}+1}\right\} 
q^{\Delta_{n,m}^{+}}\bar{q}^{{\Delta}_{n,m}^{-}}\end{eqnarray*}
where 
$$ \Delta_{n,m}^{\pm}=\frac12 \left( \frac{n}{R}\pm \frac{m}{2}R
\right)^2$$
 are the conformal weights of the Gaussian part of the CFT,
$\eta(q)$ is the Dedekind function, and $q=e^{2\pi i\tau}$,
$\tau$ being the modular parameter.

This analytical result agrees with the earlier conjecture of Alcaraz and Martins based on
numerical investigation of the Bethe Ansatz equations of the spin-1 XXZ chain \cite{23}. 

To close this section we remark that the results for the conformal weights (\ref{Dpm}-\ref{R})
are analytic in the anisotropy parameter $\gamma$ in the entire regime $[0,\pi/2]$ 
although the Bethe ansatz patterns for the excited states look qualitatively different
 for the repulsive and the attractive regimes. The situation is quite similar to that
of the spin-1/2 chain. There, the same derivation that is successful in the
repulsive regime is applicable in the attractive regime. The key to this is
the large freedom in the choice of the integration contour $\Gamma$ and the
possibility to write down non-linear integral equations that are algebraically
identical to those for the ground state but using deformed contours. In this
approach there are no additional "driving terms" in the NLIEs and the
conformal charges are given in terms of dilogarithm functions with
"non-standard" integration contours \cite{12b}. 
The analyticity of conformal weights in the spectral parameter indicates that
this approach can probably be extended for higher spin cases too.

\section{Ordinary and virtual special objects}

In this section we would like to shed more light on the appearance of virtual special objects in the NLIE, 
therefore we will consider the ground state as simplest example.
The ground state of the model is formed by $N/2$ quasi 2-strings with quite small deviations
from $\pm \frac{\pi}{2}$ in their imaginary parts.
 First let us write down the NLIE when the contour is chosen to be a straight line running above
 all the effective 2-strings (i.e. $\Gamma_1(t)=t+i \eta_1$ with $\eta_1>\delta_j \, \,  \forall j $).
 (See figure 5.) In this case the NLIE takes the form of (\ref{nlie1})
 with trivial source terms:
 $$ g_1(\theta)=g_b(\theta)=g_y(\theta)=0.$$ 
\begin{figure}[htb]
\begin{flushleft}
\hskip 15mm
\leavevmode
\epsfxsize=160mm
\epsfbox{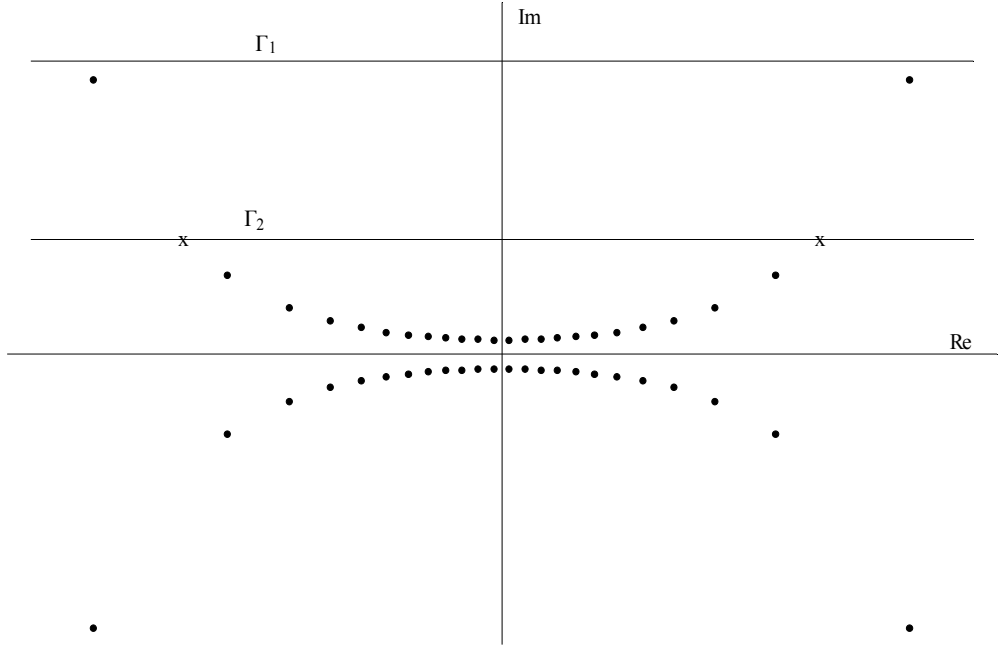}
\end{flushleft}
\caption{{\footnotesize Effective plane and 
graphical demonstration of the free choice of the integration contour in case of the ground state of the 
model. The crosses lying on the contour $\Gamma_2$ denote the positions of the virtual special objects.
}}
\label{5}
\end{figure}
In this case there are neither ordinary nor virtual special objects present.

Next we consider another straight contour $\Gamma_2(t)$ running under exactly two effective 2-strings:
(i.e. $\Gamma_2(t)=t+i \eta_2$ with $\eta_2<\delta_1=\delta_{N/2}$ and $\eta_2>\delta_j$
for all other values of $j$). (See figure 5.)
In this case the non-bulk source terms of the NLIE are not zero anymore! Two pairs of close source 
objects 
corresponding to the two effective 2-strings being above $\Gamma_2(t)$ appear as source terms
and also two virtual special objects show up in the source term of the NLIE.
In the language of the counting equation (\ref{cet2}) the appearance of the four close source objects (i.e 
$M_C=4$) entails the appearance of two virtual special objects ensuring that we are still speaking about
the ground state (i.e. $N_H=M_W=M_{sc}=S=0$). 
Then compared to the previous case the source terms of (\ref{nlie1}) are modified as
\begin{eqnarray} \label{}
g_b(\theta)&=&\chi(\theta-v_1)+\chi(\theta-\bar v_1)+\chi(\theta-v_{N/2})+\chi(\theta-\bar v_{N/2}) \\
&-&  \chi(\theta-c_1)-\chi(\theta-\bar c_1)-\chi(\theta-c_{N/2})-\chi(\theta-\bar c_{N/2}), \nonumber
\end{eqnarray}
 \begin{eqnarray} \label{}
\tilde g_y(\theta)&=&
\chi_K(\theta-v_1)+\chi_K(\theta-\bar v_1)+\chi_K(\theta-v_{N/2})+\chi_K(\theta-\bar v_{N/2}) \\
&-&  \chi_K(\theta-c_1)-\chi_K(\theta-\bar c_1)-\chi_K(\theta-c_{N/2})-\chi_K(\theta-\bar c_{N/2}), 
\nonumber
\end{eqnarray}
and $g_1(\theta)=0$. As for the notation $v_1$ and $v_{N/2}$ denotes the positions of the virtual
special objects and the set $\{ c_1,c_{N/2}, \bar c_1, \bar c_{N/2}\}=\{ \pm x_1 \pm i \delta_1\}$ denotes 
the set of close source objects. 
These source objects are subjected to quantization conditions, namely $c_1$, and $c_{N/2}$ satisfy
a condition like (\ref{eq:close}), and $v_1$, and $v_{N/2}$ have to satisfy (\ref{eq:virtspecials}). The other 
source objects
are simply complex conjugates of these ones. 
This was the simplest example how the virtual special objects enter the NLIE. Nevertheless
we have to notice that although the two different formulations of the NLIE for the ground state are 
equivalent but they are not equally practical! The second formulation containing virtual special 
objects 
is not appropriate for numerical studies, because the usual large $N$ iteration of the NLIE fails to 
converge, whilst the first formulation of the NLIE can be solved easily numerically.
Certainly when analytical calculations are considered (e.g. calculation of conformal weights) they
can be used equally well, and they give the same result.
 To summarize: the same state can be formulated in many ways by means of NLIE by the modification of 
 the contour $\Gamma(t)$. The different formulations differ from each other in 
 the number of virtual special objects occuring in the source terms. 

 The ordinary special objects defined by (\ref{sj}) according to the counting equation (\ref{cet2})
 generally accompanied by the appearance of two holes, just like in the 6-vertex model. 
 Taking a glance at the NLIE (\ref{nlie1}), ordinary special objects may pop up in those regimes where the
  derivative of the
  bulk source term $D(\theta)$ becomes small. This means that in the spin chain they can appear
  at the edges of the root distribution (i.e. $\ln N \lesssim| \mbox{Re} \, \theta| $).
  In case of alternating inhomogeneities if the inhomogeneity parameter $\Theta$ scales as $\ln N$ as
  $N \to \infty$ then ordinary special objects may emerge in the middle of the root distribution or at its
  edges. In the inhomogeneous case a nice numerical example for ordinary special objects can be found in 
appendix C of ref. \cite{6}.
We remark that when any type of special objects show up in the form of the NLIE then its numerical solution 
fails to 
converge, thus in the presence of specials one can rely on only analytical manipulations of the NLIE.
  
 \section{Summary and perspectives}  
  
In this report we extended J. Suzuki's NLIE for the attractive regime
 ($\frac{\pi}{3}<\gamma<\frac{\pi}{2}$) of the inhomogeneous 19-vertex model with alternating 
inhomogeneities.
 Then we analyzed the NLIE corresponding to the homogeneous case to describe the finite size spectrum of
  the integrable spin-1 XXZ chain. Using the NLIE we determined analytically the conformal 
  spectrum emerging in the large $N$ limit of the spin chain. Our result agreed with the one proposed
  earlier by Alcaraz and Martins based on numerical investigation of the Bethe Ansatz equations \cite{23}.
  By means of the NLIE we discussed the typical Bethe root configurations of the thermodynamic limit,
   and we worked out a method to determine the 2-string deviations for any eigenstate of the model.
   Moreover we have shown analytically for a class of excited states that,
   in accordance with earlier numerical observations \cite{22}, the holes in the sea of 2-strings are the 
positions were the 2-string deviations change sign.
   
   Finally, we would like to make a few comments on the possible further applications of the NLIE.
   First of all it is worth mentioning that recently, 
   in the framework of quantum inverse scattering method, remarkable progress has been made in the 
calculation of form factors and correlation functions of integrable spin chains \cite{4,25}.
The correlation functions are represented as complicated multiple integrals, and the form factors
of local operators are expressed as determinants of matrices, whose matrix elements are known analytic 
functions of Bethe roots of the eigenstates involved.
The multiple integral representations of correlation functions are too complicated for their explicit
analytical or numerical evaluation.    
However as it was done in \cite{25}, one can try to calculate the correlation functions numerically by 
formulating them as series of appropriate form factors. In this case the form factors are determined
by their determinant representations involving the Bethe roots of the included eigenstates.   

This method requires the knowledge of the Bethe roots for many eigenstates.
Nevertheless the solution of the Bethe Ansatz equations in higher spin
XXZ chains is not easy for large $N$ because of
the string deviations. At this point the NLIE can be useful since it can be numerically solved easily
for large $N$ and using the definitions of the auxiliary functions, all the positions
of the Bethe roots can be extracted from the numerical solution, providing a good starting point for
the numerical calculation of form factors.
 
 Another area of applications of the NLIE is the description of finite size effects in 1+1 dimensional 
integrable quantum field theories.
 It is known that the NLIE (\ref{nlie1}) with appropriately tuned inhomogeneity parameter describe
 the finite size effects in the ${\cal N}=1$ supersymmetric sine-Gordon model \cite{6,7}.
 The NLIE governing the finite size effects of the ${\cal N}=1$ supersymmetric sine-Gordon 
model in the repulsive regime was investigated in \cite{6}, and the ground state of the model with
Dirichlet boundary conditions was studied in \cite{7}.
The extension of the NLIE technique to the attractive regime, presented in this paper, enables one to 
 investigate the attractive regime of the model as well.

 \vspace{1cm}
{\tt Acknowledgements}

\noindent 
This investigation was supported by the Hungarian National Science Fund OTKA (under T049495).

\section*{Appendix A}

In section 9, in the calculation of the conformal weights,
the following dilogarithmic sum must be calculated: \begin{equation}
S_{0}=2\left\{ 
L_{+}\left[e^{3i \gamma S}\,2\,\cos(\gamma S)\right]
+L_{+}\left[e^{-3i\gamma S}\,2\,\cos(\gamma S)\right]
+L_{+}\left[\frac{\sin\left(3\gamma S \right)}{\sin\left(\gamma S \right)}\right]\right\} 
,\label{S0}\end{equation}
 where \begin{equation}
L_{+}(x)=\frac{1}{2}\int\limits _{0}^{x}\, dy\,\left\{ \frac{\ln(1+y)}{y}-\frac{\ln y}{1+y}\right\} 
.\end{equation}
 Using the dilogarithm identity of appendix B of \cite{6}, it can be easily proven that
  \begin{equation}
S_{0}=\frac{2\pi^{2}}{3}+\hat{N}_{+}\left\{ 2\pi\left|\gamma S-\pi\left[\frac{\gamma 
S}{\pi}\right]-\frac{\pi}{2}\right|-\pi^{2}\right\} -i\,\hat{N}_{+}\pi\,\ln\left(4\,\cos^{2}(\gamma 
S)\right),\label{RS0}\end{equation}
 where \begin{equation}
\hat{N}_{+}=N_{+}\,\,\mbox{mod}\,\,2\qquad 
N_{+}=\left[3\frac{\gamma S}{\pi}\right]-\left[\frac{\gamma S}{\pi}\right],\end{equation}
 and we made the choice of $\ln(-1)=i\pi$.
	 
 \section*{Appendix B}  
   
  This appendix is devoted to list the constants emerging in the kink equations (\ref{kinkb}-\ref{kinkatilde}).
  They are as follows 
\begin{equation} \label{}
C_{b\pm}=\pm 2 \chi_{\infty}(S-S^+-N_+)\pm \pi \left( \frac{N}{2}-(M_{sc}-M_{sc}^{\pm})+
\frac{N_1-N_1^{\pm}}{2} \right)+2 \pi K_W^{\pm},
\end{equation}
\begin{equation} \label{}
C_{y+}=i \pi \left(S+M_1-N_+-S^+-M_1^+ \right),
\end{equation}  
\begin{equation} \label{}
C_{y-}=i \pi \left(S^-+M_1^-+2M_{sc}^--S-M_1+N_+-2M_{sc} \right),
\end{equation} 
\begin{equation} \label{}
C_{\tilde a \pm}=\pm 2 \pi \left(S-S^\pm-N_+ \right)+2\pi N_{\tilde a \pm},
\end{equation}    
where the integers $K_W^{\pm}$ and $N_{\tilde a \pm}$ depend on the relative positions of the 
wide effective roots, and they must be determined case by case. 
  
\section*{Appendix C}
   
In this appendix the most important properties of the function $\chi_K(\theta)$ will be clarified. 
 The function $\chi_K(\theta)$ is defined as the odd primitive of the kernel $2 \pi \, K(\theta)$ and 
given by the formula:
 \begin{equation} \label{}
\chi_K(\theta)=i \, \ln \frac{\sinh\left( i \frac{\pi}{4}+\frac{\theta}{2} \right)}
{\sinh\left( i \frac{\pi}{4}-\frac{\theta}{2} \right)}, \quad
\chi_K(\theta)=-\chi_K(-\theta) \quad  \forall \theta \in {\mathbb C}.
\end{equation}
The branch cuts are chosen to run parallel to the real axis so that $\chi_K(\theta)$ be an odd real 
analytic function on the entire complex 
plane and continuous along the real axis. In this case $\chi_K(\theta)$ is not periodic  anymore with 
respect to $2 \pi i$. It is periodic only modulo $2 \pi$, i.e. the following identity holds:
$$\chi_K(\theta+2\pi i)=\chi_K(\theta)-2\pi.$$ It follows that the distance between the consecutive cuts
is $2 \pi i$ and the jump of $\chi_K(\theta)$ is equal to $-2\pi$ at each branch cut crossed from 
below to up. 
The choice of branch cuts is depicted in figure 6. 
\begin{figure}[htb]
\begin{flushleft}
\hskip 15mm
\leavevmode
\epsfxsize=160mm
\epsfbox{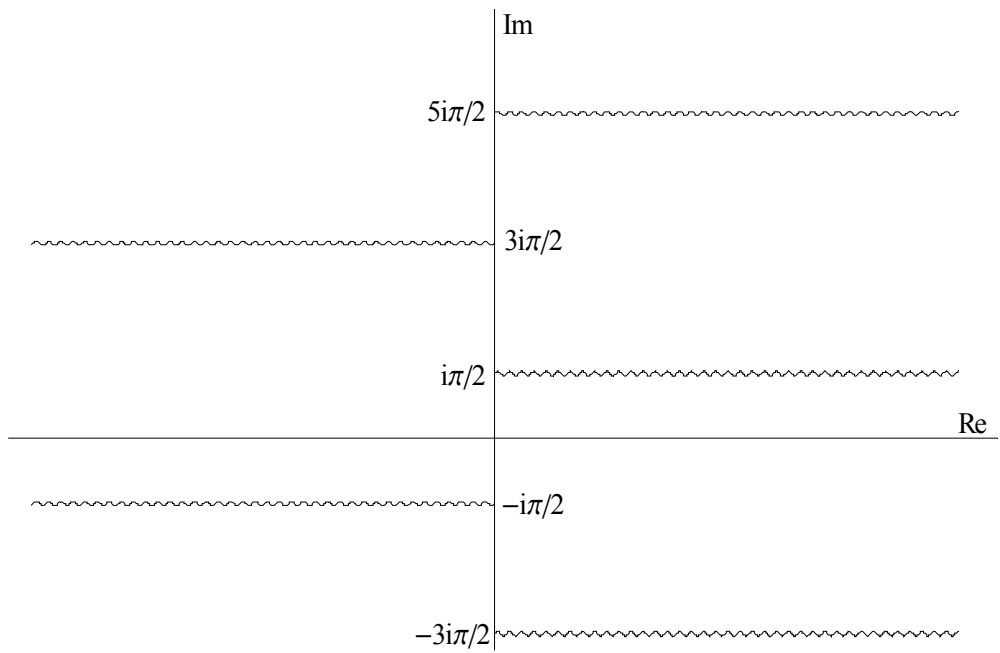}
\end{flushleft}
\caption{{\footnotesize
Locations of the branch cuts of $\chi_K(\theta).$
}}
\label{6}
\end{figure}

\end{document}